\documentclass[sigconf]{acmart}
\pdfoutput=1

\AtBeginDocument{%
  }

\copyrightyear{2024}
\acmYear{2024}
\setcopyright{acmlicensed}\acmConference[CHI '24]{Proceedings of the CHI Conference on Human Factors in Computing Systems}{May 11--16, 2024}{Honolulu, HI, USA}
\acmBooktitle{Proceedings of the CHI Conference on Human Factors in Computing Systems (CHI '24), May 11--16, 2024, Honolulu, HI, USA}
\acmDOI{10.1145/3613904.3642749}
\acmISBN{979-8-4007-0330-0/24/05}

\acmConference[CHI '24]{Proceedings of the CHI Conference on Human Factors in Computing Systems}{May 11--16, 2024}{Honolulu, HI, USA}

\acmPrice{15.00}
\acmISBN{979-8-4007-0330-0/24/05}

\usepackage{multirow}

\hypersetup{
           breaklinks=true,   
           colorlinks=true,   
        }

\begin{document}
\title[American Jews May Be Disproportionately Harmed by Intellectual Property Dispossession in Large Language Model Training]{A Canary in the AI Coal Mine: American Jews May Be Disproportionately Harmed by Intellectual Property Dispossession in Large Language Model Training}

\author{Heila Precel}
\affiliation{
    \institution{Boston University}
    \city{Boston}
    \country{United States}
}

\author{Allison McDonald}
\affiliation{
    \institution{Boston University}
    \city{Boston}
    \country{United States}
}

\author{Brent Hecht}
\affiliation{
    \institution{Northwestern University}
    \city{Evanston}
    \country{United States}
}

\author{Nicholas Vincent}
\affiliation{
    \institution{Simon Fraser University}
    \city{Burnaby}
    \country{Canada}
}

\renewcommand{\shortauthors}{Precel et al.}

\begin{abstract}
Systemic property dispossession from minority groups has often been carried out in the name of technological progress. In this paper, we identify evidence that the current paradigm of large language models (LLMs) likely continues this long history. Examining common LLM training datasets, we find that a disproportionate amount of content authored by Jewish Americans is used for training without their consent. The degree of over-representation ranges from around 2x to around 6.5x. Given that LLMs may substitute for the paid labor of those who produced their training data, they have the potential to cause even more substantial and disproportionate economic harm to Jewish Americans in the coming years. This paper focuses on Jewish Americans as a case study, but it is probable that other minority communities (e.g., Asian Americans, Hindu Americans) may be similarly affected and, most importantly, the results should likely be interpreted as a ``canary in the coal mine'' that highlights deep structural concerns about the current LLM paradigm whose harms could soon affect nearly everyone. We discuss the implications of these results for the policymakers thinking about how to regulate LLMs as well as for those in the AI field who are working to advance LLMs. Our findings stress the importance of working together towards alternative LLM paradigms that avoid both disparate impacts and widespread societal harms.
\end{abstract}


\sloppy
\maketitle

\section{Introduction}

One of the most prominent critiques of large language models (LLMs) is that they train on massive amounts of content without the consent of the authors of that content~\cite{vincent_dont_2020,vincent_chatgpt_2023, li2023dimensions,associatedpressJamesPattersonMargaret2023,rothAnotherGroupWriters2023,orlowskiInternetOriginalSin2023,atwoodMurderedMyReplica2023,senftlebenGenerativeAIAuthor2023,jiangAIArtIts2023,vincent_data_2019}. This concern is exacerbated by one of the core promises of LLMs: their ability to use patterns in their training data to substitute for the paid labor of those who created said data. People in a wide range of professions (e.g., fiction-writing and journalism) are now accusing language modeling companies of not only stealing their content (e.g., novels and news stories), but also of using this very content to put them out of a job (e.g., \cite{associatedpressJamesPattersonMargaret2023,javedMediaMogulWarns2023,smithAIStartingThreaten2024,santosChatGPTAIReplacing2024}). Indeed, the dominant approach to training LLMs has been called LLMs' ``original sin'' \cite{kevin_roose_casey_2023} and a ``property land grab'' that is ``so brazen it has unified a wide range of interests''~\cite{orlowskiInternetOriginalSin2023}. According to well-known novelist Margaret Atwood~\cite{atwoodMurderedMyReplica2023}, LLMs enable an author such as herself to be ``dispensed with---murdered by my replica...who needs the cow when the milk's free?''

Large-scale property dispossession in the name of progress---whether the property is physical or intellectual---is far from an unprecedented event, and history teaches us that it often does not occur uniformly across demographic lines (e.g., \cite{ukParliamentEnclosure, cramerHenriettaLacks}). In this paper, we seek to explore whether the ``[intellectual] property land grab''~\cite{orlowskiInternetOriginalSin2023} by LLM companies continues this historical pattern and disproportionately affects certain groups more than others. The stakes here are very high: it is not just the right to control what people can do with their content that is at risk (i.e., \textit{intellectual property dispossession harms}); the labor substitution potential of LLMs means that the ability to pursue one's chosen profession may also be seriously affected~\cite{brynjolfssonTuringTrap,acemoglu2022artificial,eloundou_gpts_2023,yilmazAIDrivenLaborSubstitution2023} (i.e., \textit{labor substitution harms}).

We focus on property dispossession from American Jews as a case study in this paper, motivated by 1) the long history
of property dispossession---including intellectual property (IP)---suffered by Jewish populations~\cite{holocaustMemorialMuseumAryanization,andreadisSeizureOfJewishIP,bannisterLetterJewishPropertyInSpain}, and 2) the contextual expertise of the author group, which can be particularly valuable given the sensitive nature of studies like this one. Importantly, however, this line of investigation is likely relevant to many other minority groups (e.g., Asian Americans, Hindu Americans) and, as we discuss below, far beyond these groups as well. Model builders' intentional decision to not provide public documentation of the data used for training contributes substantially to the difficulty in studying dispossession at scale, making a case-study based approach much more tractable at this time.

The results in this paper are clear and concerning: our findings indicate that American Jews are likely to be disproportionately affected by language models' alleged theft of intellectual property and the corresponding downstream effects on the value of their labor. Using inference methods developed in Jewish demographic literature, we find that content authored by American Jews is substantially over-represented in important LLM training datasets by between approximately 2x and approximately 6.5x. These findings raise the possibility that if the current language modeling paradigm is allowed to continue, the initial wave of economic disruption could introduce a novel and significant material bias against Jewish populations.

This paper likely has a number of significant implications for legal experts, policymakers, researchers, and the tech industry. Perhaps most urgently, our results highlight that any legal finding that current LLM training practices are allowed under some definition of ``fair use'' (or any new law along those lines; see \cite{lee_ai_2023} for an overview) could have a substantial disparate negative impact on American Jews, and probably other minority groups as well. Critically, our results should also encourage policymakers, funders, and researchers to shift resources away from the current language model paradigm and towards promising new approaches that allow for power and revenue to be distributed to content owners in a fashion that is more aligned with the value they create for LLM companies. We provide some discussion of these alternative paradigms below. 

The results in this paper also add urgency to calls (e.g., \cite{brynjolfssonTuringTrap}) to focus significantly more research and development resources on LLM applications that augment or complement human labor and significantly fewer resources on applications that substitute for human labor. This requires the creation of significant extrinsic incentives to make it more likely that this occurs (e.g., policy incentives). As we discuss below, the labor substitution harms highlighted in this paper can be significantly mitigated if we as an industry are successful in this endeavor. Many researchers think extensive labor substitution is likely (e.g., \cite{eloundou_gpts_2023, hatzius_potentially_2023, brynjolfssonTuringTrap}) and, importantly, near-total labor substitution is the explicitly stated goal of key LLM actors like OpenAI~\cite{openai_openai_2018}. However, it is possible that through changes in the sociotechnical landscape of LLMs and the technologies they power---changes that researchers and practitioners in the HCI community can help drive---some of this family of harms can be averted and even reversed.

It is important to acknowledge that this paper deals with unusually sensitive issues, the very highlighting of which in a research paper could cause harm. While we made every attempt to make evidence-backed decisions to minimize any negative effects to the Jewish community and others (e.g., from anti-Semitic actors), some risk could remain. Ultimately, the author group---of which half is Jewish---estimated that the benefits of highlighting the uncovered evidence far outweighed the potential harms of publishing it. Additionally, in developing our methodologies, we consulted with several demographers of Jewish populations, who have extensive experience handling these challenges in their work.

Finally, before continuing with Related Work, it is useful to reflect on the title we selected for the paper. It is often said that when new forms of prejudice emerge against the Jewish population, they are a ``canary in the coal mine''~\cite{freedland_antisemitism_2018, us_department_of_state_building_2020} for serious systemic issues whose harms will soon spread well beyond Jews. Applying this analogy to the findings in this paper is imperfect in some ways, e.g., we know of no evidence of the antisemitic intent that is common when the analogy is typically employed. However, the ``canary in the coal mine'' phrase does capture possibly the most important property of our results: rather than anything specific to the American Jewish community, the effects we observe here must be viewed as an additional flashing-red warning sign for the foundational flaws in the current LLM paradigm that are already affecting hundreds of millions of non-Jews as well. Indeed, without strong action, these harms may extend to nearly all participants in the economy; for example, researchers are working to use LLM techniques for robotics in ways that would create similar challenges for those who earn their living from manual labor ~\cite{Zeng2023LargeLM}. All of that said, shared harms also means shared solutions: the many promising alternatives to the current LLM paradigm that are being explored---changes that need more attention and resources---can not only remedy issues for American Jews, but can do the same for the much larger group of people outside the Jewish community who will otherwise be similarly harmed.

\section{Background}

\subsection{Broader Historical Context on Property Dispossession}
There is an extensive literature on the long history of property dispossession from marginalized groups, which we present here not to draw a direct comparison but rather to situate the current dispossession within historical context. One key teaching of this literature is that the effects of dispossession can be both tremendous and long-lasting. The nearly total dispossession of property rights---including bodily rights---suffered by African slaves in the United States led to a wealth gap that has lasted for generations since slavery ended~\cite{craemerWealthImplicationsSlavery2020}. The effect of property dispossession on indigenous peoples around the world is similarly long-lasting~\cite{carlosIndigenousNationsDevelopment2022}.

Advocates of the current LLM paradigm argue that non-consensual training on content is necessary to unlock the significant technological progress manifest in LLMs. The literature on property dispossession highlights that this type of justification is common: ``progress'', including technological progress, is often a key part of the stated justification for the seizing of property rights from marginalized groups. For example, during the English Enclosure movement, large swaths of the poor English population lost rights to farm on land they had been using for generations. New agricultural techniques were one stated reason Enclosure policies were enacted~\cite{ukParliamentEnclosure}. This justification ignored ways the ``technological dividend'' from the use of new techniques could be distributed more equitably to those who had the original property rights~\cite{fairlieShortHistoryOfEnclosure}. The dispossession of property rights that occurred in Enclosure led to decades of major riots and civil unrest~\cite{liddy_urban_2015}. An interesting precedent also comes from the story of Henrietta Lacks, an African-American woman whose cells helped create countless new innovations in healthcare but were acquired and used without her knowledge~\cite{cramerHenriettaLacks}. Her descendants recently reached a ``groundbreaking''~\cite{ozaHeLaSettlement} settlement with a major biotechnology company over their claims that the company was ``unjustly enriched'' by using her cells without her consent, an argument that some have made about the LLM industry~\cite{rothAnotherGroupWriters2023}.

Mass property dispossession from Jews has occurred for centuries. Jews in most of Eastern and Central Europe were forbidden from owning land for much of the second millennium~\cite{tellerYivoEconomicLife}, for example. Similarly, the Expulsion of Jews from Spain in 1492 was preceded by the forced liquidation of all Jewish property in that region (at significantly depressed values)~\cite{bannisterLetterJewishPropertyInSpain}. Unsurprisingly, the Nazi government in Europe in the 1930s and 1940s coupled its policies that led to the death of nearly half the Jewish population~\cite{europeJewishPop} with policies that seized nearly all Jewish property in Nazi-controlled territories~\cite{holocaustMemorialMuseumAryanization}. These policies, which are grouped under the label ``Aryanization''~\cite{holocaustMemorialMuseumAryanization}, forced Jews to either sell property at greatly reduced prices or, beginning in 1938, have it seized by the government.

The Nazi dispossession of property rights from Jews was not limited to physical property: the government also seized Jewish intellectual property. For example, in 1939, all Jews in Nazi-controlled territories had to change their middle names to ``Israel'' or ``Sara'' to increase the visibility of their Jewish identities, which had the effect of making it very difficult to renew (or apply for) any IP protections. This led to the complete reproduction of Jewish works without the consent of their authors~\cite{andreadisSeizureOfJewishIP} and even to Jewish scientists leaving their names off of patents so that the patents would be granted~\cite{barnerAryanizationExpanded}. 

Of course, no direct comparison can be made between events like the Holocaust and the IP dispossession we discuss in this paper. Indeed, we are aware of no evidence that dispossession by LLM developers intends to target the affected parties. However, the IP dispossession we identify in this paper is not without historical parallels and this context is necessary to understand the lasting effects that IP dispossession has had on affected groups throughout history.

\subsection{LLMs and Economic Harms}
For people who create content, current LLMs create two separate but compounding economic harms: 1) dispossession of intellectual property rights, and 2) the use of the dispossessed IP to substitute for their labor. Below, we discuss research that highlights the presence and significance of both of these harms. We highlight how it is useful to understand these harms separately, and while one may come to dominate in material outcomes (e.g., substitution could be the primary concern in a world in which LLMs act as labor-replacing technologies), both require addressing.

\subsubsection{IP Dispossession Harms}
\label{related:ip-dispossession-harms}
It is now public knowledge that all prominent large language models have been trained on enormous amounts of digital content---both natively digital content (e.g., Wikipedia and forum comments) and digitized documents (e.g., novels and scientific publications). With few exceptions, this is done without the consent of the creators of that content and without any form of compensation going to those creators~\cite{orlowskiInternetOriginalSin2023}. This is true both for language models that generate text (e.g., OpenAI's ChatGPT~\cite{petersNewYorkTimes2023} and Meta's LLaMA~\cite{goldsmithMichaelChabonDavid2023}) and those that generate images (e.g., Stable Diffusion~\cite{vincentGettyImagesSues2023,jiangAIArtIts2023}).

Those who make content for a living are beginning to take significant action against model builders in an attempt to reclaim some of the rights to their IP. OpenAI, Meta, and other companies that build and use LLMs are subject to a large and growing number of lawsuits, an effort that has been led in part by prominent Jewish content creators such as Michael Chabon~\cite{goldsmithMichaelChabonDavid2023} and Sarah Silverman~\cite{davisSarahSilvermanSuing2023}. Major content-creating institutions are also beginning to sue model builders; most notably, \textit{The New York Times} has sued Microsoft and OpenAI~\cite{grynbaum_times_2023,petersNewYorkTimes2023}. Legal action is far from the only avenue being used. The Authors' Guild and many others~\cite{theauthorsguildAuthorsGuildSubmits2023} are working to enact new legislation in many parts of the world to protect creator rights, and content owners are taking direct action to prevent their content from being used without their consent. For example, most prominent news websites now forbid model builders from accessing their content by editing their \texttt{robots.txt} files~\cite{bogleNewsBlockOAICrawler,petersNewYorkTimes2023}. 

There is much legal debate about how these lawsuits will be decided and in which jurisdictions. Similarly, there has been significant reporting about new laws drafted around the world to regulate language models. We discuss the implications of the results of our study for the legal debate and policy conversations in {Discussion}. It is important to note here, however, that courts and governments throughout history have blessed mass property dispossessions that are now widely considered moral abominations and have had significant negative long-term implications for both the dispossessors and the dispossessed populations. This is true of nearly all the events discussed above. As we are concerned with IP dispossession broadly, we do not focus on any specific legal jurisdiction's interpretation of what constitutes a violation of IP laws---rather, we are interested in potential harms from content use without consent. We hope to contribute to the evolving conversations around which legal doctrine(s), if any, should be used to prevent such harms.

This broad form of dispossession deprives creators of the right to choose which systems and organizations their outputs bolster, and potentially prevents their ability to receive compensation for their works. While highly related, this potential harm is distinct from the possibility that AI systems will lower the demand for future labor, which we address in the next subsection. In simple economics terms, we can think of the difference as creators losing expected compensation from content already created, or having their compensation redirected to organizations with which they are unaffiliated, versus creators losing new labor opportunities in the future. This new dispossession represents a violation of the implicit social contract that motivated people to invest time and money in training themselves; those who undertook training in a pre-LLM era had different expectations about how their work might be utilized.

\subsubsection{Labor Substitution Harms}
The use of one's creations without consent and without the ability to extract value from this use is a major harm in and of itself. However, the way that LLMs use content makes this harm potentially exponentially larger: a key stated reason for developing LLMs is so the models can learn from the content on which they are trained to ``do more and more of the work'' of content creators and cause ``the price of many kinds of labor to fall towards zero''~\cite{altmanMooresLaw}. Put another way, a core value proposition of large language models is to substitute for the paid labor of the people who create their training content. This means that the creators of content used in the model are not only fighting to get their share of the value their content is creating---they are fighting for the ability to continue working in their profession of choice at all. Many American Jews (and members of other populations that have suffered property dispossession) have grown up with the phrase ``They can't take your education away from you.'' In these ways, LLMs might very much do this, or at least may take away the ability to earn a living from an education (and other training and experience).

While the \textit{degree} of labor substitution from LLMs that will occur remains unclear, there is a growing line of work attempting to forecast the impacts of LLMs on various labor markets.  A key conclusion of this work is that professions that have contributed a lot of training content to the models may be much more affected than those that have not. For example, \citeauthor{eloundou_gpts_2023} found that workers with more formal education such as lawyers, graphic designers, and database administrators are more exposed to LLMs, which is in line with similar work forecasting the impact of ChatGPT on the labor market~\cite{zarifhonarvarEconomicsChatGPTLabor2023}.

\subsection{Dataset Documentation}
Our work aims to contribute to the ongoing discussion about data documentation, and to specifically highlight how current practices within the AI research community may be obfuscating potential harms. Many of the methodological challenges (though not all of them) presented in this paper could have been avoided with better dataset documentation, a practice that scholars have called for many times over the years~\cite{gebru2021datasheets,paullada2021data}. The field of dataset documentation provides important context for our methodological choices. Here, we give some context on this research area.

Broader interest in dataset documentation was highly influenced by work on ``Datasheets for Datasets''~\cite{gebru2021datasheets}, which proposed that every dataset be accompanied by a datasheet. This would facilitate greater transparency and enable practitioners to select more appropriate datasets for their tasks. The Datasheets concept has led to uptake in documentation practices (shaping, for instance, the Dataset track at NeurIPS\footnote{See, e.g., information about the 2023 track at \url{https://nips.cc/Conferences/2023/CallForDatasetsBenchmarks}.}). In practice, the ML community has taken tangible steps to improve data documentation for some kinds of contributions. 

However, in the space of language modeling, the massive scale (``web-scale'') of data---and the choice by the ML community to deprioritze documenting this web-scale data before using it---is a major barrier to documentation, a concern that was highlighted as early as 2021~\cite{bender2021}. In short, web-scale data is simply expensive and challenging to retroactively document, and typically lacks much in the way of structured data. As a result, much of the LLM industry suffers from ``documentation debt'', making it difficult to know even basic information about how a model was trained, e.g., what datasets were used, who authored an entry in a given dataset, etc. It is important to note that this debt is intentional and explicitly so; model builders purposely avoid releasing information about their training data~\cite{openaiGPT4TechnicalReport2023,touvron2023llama}. 

Still, efforts to document data relevant to generative AI have been undertaken. One such dataset that is open, documented in some dimensions, and widely used for LLM training is the The Pile~\cite{thePile}. This is the dataset we focused on studying, in part because the prevailing documentation practices left us few other choices. Prior work has attempted to document the BookCorpus dataset, touching on topics like copyright and acknowledging the authors that underlie the training data~\cite{bandy2021addressing}. With the purpose of identifying the authors whose work was used in training Meta's LLaMA, a journalist recently processed and identified over 170,000 books contained in the Books3 dataset, finding that the majority were still under copyright and were published in the last two decades~\cite{atlanticReisner2023}.
There is also ongoing work aimed at urgently remedying the dataset documentation crisis in the context of LLMs. For instance, ``The Data Provenance Project'' has compiled metadata on popular LLM training datasets~\cite{longpredata}, revealing a large amount of prevailing ambiguity and the need for policy guidance.

The lack of consistent documentation in web-scale training datasets is especially relevant to this research because author attribution is necessary for estimating group-level dispossession. As a result, we were forced to exclude some datasets (most saliently, all web crawl data) from our analysis. 

\subsection{Contributor Attribution}

\subsubsection{Challenges of Ethnic Identification}
\label{related:challenges-of-ethnic-id}

Compounding the data documentation issue for this research is the fundamental difficulty of ethnic identification, especially for small ethnic and ethno-religious populations. Ethnic boundaries are often fluid and hard to define, and are only sometimes captured in demographic surveys. Assigning labels to individuals without directly surveying them is itself fraught and will inevitably miscategorize some members of a population, but a true survey of all group members is prohibitively expensive in both time and resources. Yet, it is critical for communities to understand the needs and challenges facing their members, and formal population measurements can be the gateway to official recognition and institutional support~\cite{mateosClassifyingEthnicityThroughNames2014}. As a result, groups have developed a variety of alternate strategies for counting and surveying their members.

In this paper, we focus on the American Jewish community, a small ethnoreligious group (about 1.8--2.4\% of the U.S.\ population~\cite{ajppBrandeisReport, pew2020Report, sarnaAmericanJewishPopulationEstimates}) with a robust literature of community studies at local and federal levels (e.g., \cite{himmelfarbSamplingByEthnicSurnames, njps2000Data, sheskinMiami1998, pew2013Report, markerJewishCommunityStudies21stCentury, allAJYB}). These studies are designed to capture the size, character, demographic profile, and needs of U.S.\ Jews and synthesize findings into actionable insights for the Jewish community~\cite{boxerAllPoliticsIsLocal2016}. However, the U.S.\ Census doesn't collect information on religion and the Jewish community is small enough to make identifying a representative sample via Random Digit Dialing (RDD) or other common random surveying approaches extremely costly~\cite{ajppBrandeisReport}. As a result, Jewish demographic studies have relied on probabilistic methods to identify a representative sample and extrapolate findings to communities at large.\footnote{Defining membership in the Jewish community is a complex topic, discussion of which is beyond the scope of this paper. We rely on definitions from Brandeis's AJPP Report~\cite{ajppBrandeisReport}, the AJYB population counts~\cite{allAJYB}, and Pew Research Center's Jewish American studies~\cite{pew2013Report, pew2020Report}. For an overview of how these definitions were developed, see pages 3-6 of \cite{ajppBrandeisReport}.}

\subsubsection{Distinctive Jewish Names (DJNs)}
\label{related:djn}

One method developed by the American Jewish community for in-group surveying is the Distinctive Jewish Names (DJN) frame. In this method, lists of potential survey respondents (whether via landline RDD, cellphone dialing, or otherwise) are filtered to candidates with a surname that is distinctively Jewish: that is, both common in the Jewish community and largely unique to Jews. This increases the chances that a respondent will be Jewish, thus potentially lowering survey costs. The DJN frame has a rich history: it was initially proposed in the 1940s by Kohs ~\cite{himmelfarbSamplingByEthnicSurnames} and has since been used in numerous community studies through to the present day~\cite{allAJYB}. In some cases, DJN-based lists were used alongside other sampling methods, primarily RDD; in others, they were the primary or only sampling frame. Some studies were designed to actually survey community members while others estimated a Jewish population count in a particular context based on DJNs in membership lists or phone books. See the ``United States Jewish Population'' chapters of the American Jewish Yearbook series for a detailed account of such studies~\cite{allAJYB}.

In this study, we use DJNs as our primary form of analysis. We estimate the size of a Jewish population from a total list of names, using content authors identified from model training data as our population. Here, we provide a brief overview of our rationale before elaborating on the method itself in Section~\ref{methods}. For a more detailed overview of the literature on DJNs and the frame's limitations, see Appendix~\ref{app:djns}. 

In general, a DJN-only approach is recommended for identifying the size of a Jewish population a) only as a rough estimate, and b) only when one already has prior knowledge of Jewish population estimates for the area of interest. Both of these are true for our study: we are reporting order of magnitude estimates at the national scale, for which we have a number of high-authority population estimates~\cite{pew2020Report, ajppBrandeisReport}.

Additionally, the DJN-only approach to population estimation assumes that Jews in the sample with DJNs are representative of those without; in this case, that Jews with DJNs are no more likely than Jews without DJNs to produce content that appears in LLM datasets. There is some prior evidence to suggest that DJN samples do not show significant differences in income~\cite{cohenDeficientIfNotDistorted2016, sheskinGoodPractices2016} and education~\cite{himmelfarbSamplingByEthnicSurnames}---two potential proxies for IP generation---as compared to the general Jewish population. The greatest disparities are around Jewish religious knowledge and engagement with Jewish life. Absent a compelling hypothesis for why DJN samples would significantly differ from the general population on occupation or writing output, one can have reasonable confidence that our estimates provide order-of-magnitude bounds.

Ultimately, the research in this paper assumes that characterizing the nature of IP dispossession by LLMs with respect to potential impact on the Jewish community is important enough to use the best method available, even if it won't provide a highly-precise point estimate. Our analysis accounts for this by making reasonable assumptions that lead to upper and lower bound estimates with robustness checks to ensure order-of-magnitude correctness. Details on these assumptions are in Section~\ref{methods:assessing-jewish-authorship}.

\subsection{New LLM Paradigms}
Researchers have begun to explore new LLM paradigms that seek to minimize the ethical, legal, and other risks of the current approach which depends on uncompensated access to vast amounts of content used without the owner's consent. These explorations are a burgeoning area of research.

One promising direction emerges out of the retrieval augmentation and enhancement literature (e.g., \cite{zamaniRetrievalEnhancedMachineLearning2022}). These techniques allow a small base model, trained on either public domain or full-consent content~\cite{minSILOLanguageModels2023a} and combined with document retrieval techniques at runtime to dynamically generate output (e.g., \cite{lewisRetrievalAugmentedGenerationKnowledgeIntensive2020}). This approach allows enough transparency and control for individual content owners to be able to make decisions about where they want their data to end up, and bargain for specific contracts that pass value back to content owners based on usage or related metrics.

Recent work has begun to explore exactly how markets operated with carefully designed sharing incentives (e.g., ``data consortia'' in which multiple organizations pool their content together) might work in practice~\cite{castrofernandezDataSharingMarketsModel2023,xiaDataStationDelegated2022}. Scaling up support for this kind of data sharing is another way to shift LLMs towards sharing some portion of their economic winnings with content creators. Finding ways to make participation appealing to LLM developers could be an effective way to work towards an alternative LLM data paradigm.

Another idea that has been the subject of early discussions about the economic impacts of AI and automation is to implement some kind of broad ``data dividend''~\cite{vincent_sharing_2023}, through which the profits from AI technologies are shared with training data creators. A criticism against this idea is that a very broad remuneration system might hurt incentives for the creation of new content (compared to e.g. new content markets). However, this option could be complementary to other approaches: because many groups have content and IP in the training sets for web-scale models, there is a strong argument for at least some degree of broad remuneration.

As noted above, the individual viability of these alternative paradigms may change suddenly if certain legal decisions are reached (e.g., if ``fair use''-based training is broadly supported, or broadly banned) or if new regulations are passed. Thus, navigating the sea of possible paradigms will require the consideration of all possible options, and ideally will include experiments with the proposals listed here and more.
\section{Methods}
\label{methods}
The key methodological challenge of this research is figuring out how a group concerned about disproportionate IP dispossession and labor substitution in the wake of language models might go about quantifying the costs it is likely to face. Addressing this question as it pertains to Jewish Americans involves dealing with a large amount of unavoidable noise.

In this section, we discuss how we sought to reduce the amount of noise in our estimates to a minimum. We describe our methods in two parts: first, we describe the LLM training datasets we use in our analysis and the process we use to collect metadata; and second, we present our strategy for assessing Jewish authorship.

\subsection{Datasets}
Here, we describe the datasets we used, starting with the rationale underlying dataset selection. 

We wanted to analyze datasets that would give a reasonable estimate of the overall relative magnitude of intellectual property dispossession faced by Jewish authors. We assume that an IP dispossession event occurs each time an author's work is included in a training set without that author's explicit consent. For this analysis, a single work equals a single document (e.g., scientific paper, law paper, code repository, book) and works with \textit{n} authors count for \textit{n} IP dispossession events. We note that, as discussed in Section~\ref{related:ip-dispossession-harms}, dispossession is separate from (but related to) copyright, and copyright practices vary between the types of documents.

If we had metadata describing the author(s) of every document in a web-scale training dataset and the group identifications of those authors, we could very quickly identify which groups are most impacted. For instance, if a firm were able to scrape the entire web and performed no filtering, the degree of disparate impact would map directly to group-level differences in web content creation. Whichever group published the most documents on the web would see the most IP dispossession in absolute and relative terms. However, in practice the entire web is not used directly for training (see, e.g., \cite{openaiGPT4TechnicalReport2023}). Major firms use various filtering processes to select only some works for use in training data. The processes used by most major firms are currently proprietary~\cite{openaiGPT4TechnicalReport2023}, creating a significant barrier to analyses like those in this paper.

However, there exist ``open'' LLMs that, because they release their training data, implicitly reveal their filtering methods. One very popular dataset used by these LLMs---and the one we analyze in this research---is the Pile~\cite{thePile}, curated by EleutherAI. The Pile includes a variety of high-quality data sources that are largely associated with specific professions and institutionalized platforms (e.g., ArXiv, GitHub, FreeLaw). Many of these include some level of author attribution for individual works.

Open models, such as those trained using the Pile, are seeing substantial gains and catching up in performance with private offerings (see, e.g., comparison between GPT-NeoX-20B and GPT-3 \cite{blackGPTNeoX20BOpenSourceAutoregressive2022}, as well as the performance of the LLama models \cite{touvron2023llama}). This suggests that the data quality filtering and weighting used in the Pile is somewhat comparable to private filtering strategies, and that by studying this heuristically filtered (i.e., carefully selected) data, we can make claims that generalize reasonably well to LLMs as a class of technology. In other words, we expect the general practice of studying and documenting ``open'' web-scale training data to provide insights that apply to commercial LLM products. While it is unknown if Meta's ``open'' Llama 2 model directly used the Pile as details about pretraining are omitted~\cite{touvron2023llama}, it seems likely (the original LLaMA paper did report using The Pile~\cite{touvronLLaMAOpenEfficient2023}).

We focused in our analysis on high quality subsets of the Pile that clearly map to content that is unambiguously composed of individual pieces of literary, scientific, artistic, and/or professional works that are typically subject to intellectual property governance and norms. By estimating the number of (document $\times$ author) pairs present in the Pile, we get a general assessment of the magnitude of IP dispossession and potential labor substitution faced by authors whose works are in the Pile.

\subsubsection{Dataset Selection and Curation}
Below, we describe the specific datasets from the Pile that we studied in order to work towards an estimate of the relative exposure of Jewish Americans to IP dispossession. We also describe some of the dataset-specific processing steps we followed, as well as our data processing pipeline at a high level.

The Pile consists of a set of plaintext documents derived from a set of datasets with minimal additional processing. At time of writing, it is hosted by EleutherAI~\cite{eleutherAIHomepage}, and many of the subsets are also available via original sources. For our analysis, we selected five Pile components ordered by weight, which the Pile documentation defines as ``percentage of bytes in the final dataset occupied by each dataset''~\cite{thePile}: PubMed Central, Books3, ArXiv, GitHub, and FreeLaw. We excluded web scrape components (Pile-CC and OpenWebText2) because we were unable to identify usable author metadata from web scrape data and re-linking this metadata would be prohibitively difficult. Overall, the five components used in our analysis total 49.14\% of the final Pile dataset by weight~\cite{thePile}. We processed each of the datasets as follows: \\

\begin{enumerate}
\item \textbf{PubMed Central}. We downloaded the PubMed Open Access Subset directly from NCBI~\cite{pmcData}. We used the June 2023 baseline bulk files for our analysis and PubMed Parser~\cite{pubmed-parser} to parse metadata (including author surname) for each article dropping 5 unreadable files (out of 3,529,109 total).
    \\
\item \textbf{Books3}. 
We obtained the Books3 metadata from the website of the creator of the Books3 dataset~\cite{archiveBooks3Metadata}. The json file contained fields such as title, authors, publication details, and description. Any errors in the author name field will result in an undercount of Jewish authors, as mislabeled documents will still be included in the denominator, regardless of whether the authors have DJNs.
It is important to note that Books3 is a particularly controversial component of the Pile; large model builders are currently being sued by book authors for their use of Books3 (e.g., \cite{rothAnotherGroupWriters2023}).
\\

\item \textbf{ArXiv}
We downloaded ArXiv metadata from the official ArXiv Kaggle dataset maintained by Cornell University~\cite{arxiv-public-datasets}. Data was downloaded on July 23rd, 2023. We parsed author names using Clement et al.'s ArXiv data tools~\cite{arxiv-public-datasets, arxiv-public-datasets-Paper}.
\\

\item \textbf{GitHub}
We downloaded the list of GitHub repositories used in the Pile from EleutherAI's GitHub Downloader~\cite{github-downloader} and used the GitHub API~\cite{githubAPI} to collect author names for a uniformly distributed random sample of $\sim$5\% (9,980) of repositories. We used a random sample because limitations set by the GitHub API prevented us from downloading data for all repositories. The ``Name'' field is often unpopulated on GitHub profiles, and even when populated is not standardized. We parsed this field by selecting the most commonly occurring pattern (<First Name> <Last Name>).
\\

\item \textbf{FreeLaw}
We downloaded FreeLaw's CourtListener Opinion and People datasets from FreeLaw's bulk data files~\cite{courtListener-API} on May 31, 2023. We joined these datasets based on the Author ID column and filtered out opinions with no authors.
\end{enumerate}

\subsection{Assessing Jewish Authorship}
Here, we describe the process we used to identify Jewish authors in the datasets described above and produce an estimate of the relative magnitude of IP dispossession experienced by Jewish Americans.

\label{methods:assessing-jewish-authorship}
\subsubsection{Name Classification}
\label{methods:name-classification}
The DJN list we used for this analysis is a 35-name list that has been used in studies from the 1940s until the present day with very few changes, and has consistently maintained a relative proportion of roughly $\sim$10--12\% of Jews with DJNs to Jews as a whole in large American Jewish communities \cite{himmelfarbFurtherComments1876, sheskinMiami1998, sheskinCollegeCampus2013}. It was also designed to have very high precision: i.e., people with these surnames are highly likely to be Jewish. See Section~\ref{related:djn} for additional details on these numbers.

\subsubsection{Data Processing Procedure}
\label{methods:adt-data-processing}
We conducted several levels of additional processing that culminated in calculating an estimated percent Jewish authorship via a DJN-based frame. At a high level, our data processing consisted of five steps. We follow these steps for each of our five Pile subsets.

\begin{enumerate}
    \item Extract author names
    \item Match DJNs to (document $\times$ name) pairs and calculate percentage of pairs with DJNs
    \item Adjust estimated percentage to account for non-Jews with DJNs
    \item Adjust estimated percentage to account for Jews without DJNs
    \item Estimate an expected percentage of U.S.\ Jewish authors per dataset (i.e., the number we would expect to see if U.S.\ Jews were proportionally represented) and compare it to the observed percentage
\end{enumerate}

Below, we explain each of these steps in more detail. \\

\textbf{Extract author names.} First, we attempted to extract the authors of each document in each dataset to create a list of (document $\times$ name) pairs. Using (document $\times$ name) as our unit of measurement allows us to center authors as those experiencing harm: each pair represents one instance of a given author's work being used in LLM training data.

\textbf{Match DJNs to (document $\times$ name) pairs and calculate percentage of pairs with DJNs.} We use our DJN list as a filter on author names to produce a subset of DJN-matched documents. We calculate the percentage of resulting (document $\times$ name) pairs for which the name is a DJN.

\textbf{Adjust estimated percentage to account for non-Jews with DJNs.} To account for the potential that DJN matching may identify some people who are not Jewish, we rely on false positive estimates from Himmelfarb and colleagues~\cite{himmelfarbSamplingByEthnicSurnames} (``about 90--92\% of these names are Jewish''), Rosenwaike's analysis of leading Jewish surnames, which includes estimates of 76.6--95.1\% (m=88.7\%) precision for 15 of the 35 names~\cite{rosenwaikeLeadingSurnames}, and Phillips' 91.8\% Boston-area estimate \cite{phillipsNumberingTHeJews2007}.

Because these are all approximations, we use a range of 80--90\% for our analysis. In other words: if our method found 1000 (document $\times$ name) pairs that matched DJNs, we assumed this represented between 800-900 Jewish authors.

\textbf{Adjust estimated percentage to account for Jews without DJNs.} For this step, we again refer to~\cite{himmelfarbSamplingByEthnicSurnames, lazerwitzSomeCommentsOnDJNs1986, phillipsNumberingTHeJews2007}, which estimate that Jews with names from the DJN list we used comprise $\sim$10\%--12\% of the U.S.\ Jewish population. We validated this number by calculating the percentage of people included in the 2010 U.S.\ census \cite{us-census-list} with one of the DJNs, accounting for false positives as above, and comparing it to estimates of the Jewish U.S.\ population in 2010~\cite{pew2013Report, sarnaAmericanJewishPopulationEstimates}.\footnote{We use the following equation, with a precision of 85\% and a Jewish population of 1.8-2.2\%: $\frac{\text{\# of DJNs in the population } * \text{ precision}}{\text{U.S. population } * \text{ \% of Jews in the U.S.}}$}

We used the resulting figure of 9.15--11.18\% for our analysis. Following from the above example, if our method finds 1000 DJN pairs (adjusted in step four to a range of 800-900 Jewish authors), this means we'd extrapolate in step four to a lower bound of $800 * \frac{1}{0.1118} = 7156$ Jewish authors total.

We note that both of these adjustment steps simply involve multiplying DJN percentage values, so their order does not matter. The ability to perform these adjustment steps with only multiplication rests on two distributional assumptions: 
\begin{itemize}
    \item That the distribution of names among Jewish Americans who contributed to LLM training data is roughly the same as the population of \citeauthor{himmelfarbSamplingByEthnicSurnames} \cite{himmelfarbSamplingByEthnicSurnames} and \citeauthor{rosenwaikeLeadingSurnames}'s \cite{rosenwaikeLeadingSurnames} respective studies.
    \item That there is no difference in job category representation and propensity to contribute IP to LLM training data between Jewish Americans with DJNs and those without (i.e., for each DJN document-name pair, there is a proportionate number of document-name pairs that would be attributable to Jewish Americans without DJNs upon deep investigation).
\end{itemize}

\textbf{Estimate an expected percentage of U.S.\ Jewish authors per dataset.} Finally, the fifth step is to contextualize these estimated percentages in terms of relative dispossession magnitude. We want to know how the observed amount of IP from Jewish American authors compares to the expected amount of IP from Jewish American authors if LLM operators were to representatively sample works from the whole population. Because we focus here on the relative representation of Jewish American authors, we introduce two new factors we must account for: changing demographics over time and how much data in each dataset comes from American authors.

Some of the datasets we investigate represent content that was produced over many decades. In Appendix~\ref{app:dataset-age-of-freelaw}, we explored whether our estimates would change if we accounted for changes in the Jewish population over time. We tested this using FreeLaw---the dataset with by far the largest time window---and found minimal impact on our results.

To account for country-level distribution of training data, we estimated the percentage of documents from each dataset published in the U.S.\ and account for this in calculating our ``expected percentage'' of U.S.\ Jewish content. One can think of these numbers as checking how overrepresented U.S.\ authors might be in general in order to correctly calculate the percentage of American Jews.

    \begin{itemize}
        \item \textbf{Free Law:} 100\% published in the U.S., as CourtListener only indexes U.S.\ opinions.
        \item \textbf{Books3:} We do not have data on what percent of Books3 is international. As a result, we act as if 100\% of the dataset were published in the U.S.\ to intentionally use a conservative lower-bound estimate of the actual figure, even at the top end of our ranges.
        \item \textbf{GitHub:} 24.6\% published in the U.S., based on data from \cite{wachsGeographyOfGitHub}.
        \item \textbf{PubMed Central:} 28.6\% published in the U.S., based on our own estimation (details in Appendix~\ref{app:pubmed-central}).
        \item \textbf{ArXiv:} 27.3\% published in the U.S., based on our own estimation (details in Appendix~\ref{app:arxiv}).
    \end{itemize}

We note that there is another source of uncertainty relating to geography: some of the authors with DJNs could be non-American Jewish authors. We don't expect this to change our results substantially because we expect surname distributions and spellings to be different in other countries, and around 40\% of the world Jewish population lives in the U.S.

\subsection{Ethical Considerations and Broader Impacts}
The approaches we use in this work---and indeed the choice to conduct the research at all---were carefully weighed against the methodological challenges and ethical questions of doing so. Specifically, many of the methodological choices above were designed in part to minimize potential harm to the Jewish community and other stakeholders.

First, we considered the risks and challenges of labeling authors as Jewish. Identity inference from attributes like names is not new, but it is controversial~\cite{dignazio2020what}. One area with significant prior work and critique is the realm of gender: multiple studies in HCI have documented the harms of gender detection and recognition systems, from individual harms of misgendering~\cite{hamidi2018gender} to societal harms from operationalizing reductive and exclusionary definitions of gender~\cite{keyes2018misgendering}. As in our study, it is not always possible to get affirmative self-identification for attributes; yet knowing the demographic distribution in a dataset can be a critical aspect of evaluating the impact of systems.

We took a number of actions to mitigate common risks in identity inference. The list of DJNs and other methodological approaches we used were selected in consultation with Jewish demographers, who are experts at navigating the fraught ethical choices surrounding the inference of Jewish identity. One alternative approach we considered but ultimately rejected was deploying a large-scale survey to all of the authors of works in the Pile we could identify to inquire as to their claimed ethno-religious identity. However, surveys like these targeting the Jewish population are known to be very difficult to execute due to the small size of the Jewish population. Also, deploying a survey asking about Jewish identity of course has its own ethical considerations, and even if successful, would not remove the significant noise present due to the LLM community's poor documentation practices.

Additionally, as we are attempting to calculate the \textit{proportion} of Jewish authors in these datasets, our method does not require that any specific author be labeled Jewish with 100\% accuracy, thus minimizing the harm of misclassifying specific individuals (either as Jewish or as non-Jewish).

The second major ethical consideration for our study was the risk that our findings would be used to harm the Jewish community should they be leveraged by anti-Semitic actors, which we further discuss in Appendix~\ref{app:djns}.
Ultimately, we decided that the consequences of not doing the work were greater than the potential harms, a decision-making process that was led by the Jewish members of the authorship group. We stress that IP dispossession is happening regardless of how well it is documented, and that it will continue to happen until the broader LLM community takes measures to change the approaches in their work.

\subsection{Methodological Limitations}
As described above, our method required us to make several careful assumptions in order to obtain reasonable, bounded estimations for the proportion of Jewish Americans in each of these datasets.
We note that these estimation approaches are somewhat atypical in the HCI community, which often operates with better documented data. When deciding to do this work, we took inspiration from the carbon impact estimation research literature, which also operates in a very high-stakes domain and has to use a wide variety of estimation techniques with large informal error bounds~\cite{taoTrendVirtualHybrid2021}. That literature has shown that if the research question is important enough, estimates with somewhat wide ranges and important qualifiers can greatly assist with decision-making towards critical goals~\cite{taoTrendVirtualHybrid2021}.
\section{Results}
\label{results}

\begin{table*}[ht]
    \centering
    \begin{tabular*}{0.72\textwidth}{lllll}
    \hline
    \textbf{Dataset} & 
    \textbf{\begin{tabular}[c]{l@{}}\% IP with \\DJN author\end{tabular}} &
    \textbf{\begin{tabular}[c]{@{}l@{}}\% IP with \\ U.S. Jewish\\ author\end{tabular}} &
    \textbf{\begin{tabular}[c]{@{}l@{}}\% Expected \\ IP with U.S. \\ Jewish author\end{tabular}} & 
    \textbf{\begin{tabular}[c]{@{}l@{}}Relative Dispossession \\ Magnitude \end{tabular}} \\ \hline
    
    \textbf{PubMed Central} & 0.19         & 1.39-1.91         & 0.5-0.7         & \textbf{2.02-3.71} X         \\
    \textbf{Books3}         & 0.98  & 7.01-9.64  & 1.8-2.4         & \textbf{2.92-5.36} X  \\
    \textbf{ArXiv}          & 0.28         & 2.01-2.77         & 0.5-0.7         & \textbf{3.07-5.63} X         \\
    \textbf{GitHub}         & 0.29         & 2.08-2.86         & 0.4-0.6         & \textbf{3.53-6.46} X         \\
    \textbf{FreeLaw}        & 0.93         & 6.65-9.14         & 1.8-2.4         & \textbf{2.77-5.08} X         \\
                            &              &                   &                                                \\
    \textbf{Total}          & 0.54  & 3.83-5.26  & 1.0-1.3  & \textbf{2.86-5.25} X  \\
    \textbf{Weighted Total} & 0.37  & 2.63-3.61  & 0.8-1.0  & \textbf{2.46-4.51} X    \\ \\     
    \end{tabular*}
    \caption{From left to right: (1) percentage of dataset (documents $\times$ authors) with DJNs; (2) estimated percentage of dataset (documents $\times$ authors) with U.S. Jewish authors; (3) expected percentage of dataset (documents $\times$ authors) with U.S. Jewish authors; (4) relative dispossession magnitude.}
    \label{fig:PIPDE-final-estimates}
\end{table*}    

Using the method described above, we calculated \textit{relative dispossession magnitude}---a ratio of observed to expected numbers of documents written by U.S.\ Jews in the dataset.

\begin{gather*}
\text{Relative Dispossession Magnitude} = \\
\frac{\text{\% U.S.-Jewish authored documents in dataset}}{\text{Expected \% U.S.-Jewish authored documents in dataset}}
\end{gather*}

\vspace{5.75pt}
We first calculated lower and upper bounds for the relative dispossession magnitude of each individual dataset based on the lower-bound and upper-bound estimation techniques discussed above. Then, we calculated two averages: a total relative dispossession magnitude (the mean across datasets with each dataset weighted equally) and a weighted total relative dispossession magnitude (the mean across datasets with each dataset weighted by number of documents $\times$ document size). In other words, weighted total relative dispossession magnitude reflects the total overrepresentation of American Jewish authorship accounting for the size of each individual dataset and the length of its average documents.

Looking at the first column in Table~\ref{fig:PIPDE-final-estimates} (which is not limited to U.S.\ documents), we see that the percent of (document $\times$ name) pairs whose authors have DJNs---who represent a small fraction of Jewish authors---is already greater in almost every case than the percent of Jews in the world (0.19--0.28\%) \cite{dellaPergolaAJYB2021, unGlobalPopulation}. Although we have less certainty about world statistics as many of our variables are designed to focus on the U.S.\ Jewish population, this is strong early support for the hypothesis that American Jews are over-represented in these datasets and suggests that this is true of Jewish authorship globally.

In the second column, we have lower and upper bounds for the percent of IP dispossession events---(document $\times$ name) pairs---from U.S.\ Jewish authorship. As noted above, we consider a range of parameter values to account for some of the uncertainty introduced by our methods. Our estimate here is parameterized by \textit{precision} (how unique are the DJNs to the Jewish population) and \textit{coverage} (how much of the Jewish population do the DJNs represent). Our lower bound uses lowest precision / highest coverage estimates; our upper bound uses highest precision / lowest coverage. 

Table~\ref{fig:PIPDE-parameter-table} shows the parameters we use for each estimate: precision ranges from 80\% to 90\%, coverage from 9.15--11.18\%, and percent of the U.S.\ that is Jewish from 1.8\%--2.4\% (see Section~\ref{methods:adt-data-processing} for details). Critically, these ranges are not uncertainty ranges: they are assumption-based and indicate a range of reasonable possibilities for a parameter in our equation. Our lower bound is the lowest possible estimate given our assumptions; our upper bound is the highest. In Appendix \ref{app:how-wrong}, we include a robustness check in which we consider the most extreme possibilities for each parameter that would still demonstrate a relative dispossession magnitude $>$ 1.

We found that U.S.\ Jewish authorship ranged, per dataset, between 1.39--9.64\%. As expected, the more U.S.-centric sources see higher percentages of U.S.\ Jewish-authored documents (6.7--9.1\% for FreeLaw; 7.0--9.64\% for Books3) while less U.S.-centric sources see lower percentages (PubMed Central: 1.4--1.9\%, ArXiv: 2.0--2.8\%, GitHub: 2.1--2.9\%). 

\begin{table}[H]
    \centering
    \begin{tabular}{lll}
    \hline
    \textbf{Parameter} & 
    \textbf{Lower bound} &
    \textbf{Upper bound} \\
    \hline
    
    \% precision of DJNs                 & 80\%     & 90\%  \\
    \% coverage of DJNs                  & 11.18\%  & 9.15\%  \\
    \% of US population   & 2.4\%    & 1.8\%  \\
    \hspace{1em} that is Jewish & & \\ \\
    \end{tabular}
    \caption{Parameters used in estimation calculation. From left to right: (1) name of parameter; (2) value used for lower bound estimations in Table~\ref{fig:PIPDE-final-estimates}; (3) value used for upper bound estimations in Table~\ref{fig:PIPDE-final-estimates}.}
    \label{fig:PIPDE-parameter-table}
\end{table} 

The final column in Table~\ref{fig:PIPDE-final-estimates} shows the amount of dispossession experienced by U.S.\ Jewish authors relative to U.S.\ content producers more generally, i.e., the numbers we are most interested in for the purposes of this paper. The results in this column clearly show a structural bias against U.S.\ Jews across all datasets: the lowest lower-bound dispossession magnitude we observed was 2.02, meaning that U.S.\ Jewish suffer double the dispossession of the U.S.\ population as a whole at the very minimum (across the datasets we considered). The highest upper-bound magnitude was 6.46, which corresponds to over six times more dispossession than the general population. This table presents a strong argument that, at least with respect to U.S.\ content, LLMs rely disproportionately on Jewish American intellectual property obtained without the creator's consent, and do so extensively. 

\section{Discussion}
\label{discussion}
Our results indicate there is very real risk that Jewish Americans may see substantial and disproportionate economic harms as LLM-based technologies are deployed more widely. Below, we discuss the implications of these results for a number of key discussions happening around AI law, regulation, and practice. We also highlight key areas of future work.

\subsection{Implications for Legal and Policy Discussions}
The findings above have important implications for the rapidly-developing legal and policy debates surrounding language models. The introduction of new structural material biases against minorities has not been broadly considered in these debates, and our results suggest they very much should be. 

With regards to developments in the legal sphere, many dimensions of LLM training practices are being examined by courts in different jurisdictions, e.g., ``fair use,'' privacy rights, publicity rights, labor law, contract law and many others \cite{congressionallegalserviceGenerativeArtificialIntelligence2023, samuelsonGenerativeAIMeets2023, lee_generative_2023}. However, our results suggest that disparate material impact suffered by protected groups is another dimension that needs to be explored. It is clear, for instance, that any U.S.\ decision that current LLM training practices constitute ``fair use'' could introduce a significant new structural bias that disproportionately harms American Jews---and likely other minority groups---in the short term. As discussed in the Introduction, should these disparate harms play out, they are likely to represent a ``first wave'' of harms, with almost all participants in the economy eventually being affected.

The evidence above also suggests that policymakers should more deeply consider structural bias against minority groups in the discussions about language models. Lawmakers in the United States considering encoding current LLM content usage into law must wrestle with the new systemic biases against American Jews they would be creating. Similarly, those working on efforts to strengthen the rights of content owners and producers (e.g., \cite{knibbs_congress_2024}) may consider our findings to be an additional reason to push forward in that direction. Our results also highlight the importance of agency decision-makers, regulators, and policymakers shifting the growing amounts of public funding for LLM research towards the many promising approaches that do not create the structural biases of current approaches. There are a number of other reasons that have been identified in the literature to do so, e.g., the potential of the current paradigm to substantially decrease the material welfare of the general public~\cite{brynjolfssonTuringTrap, altmanMooresLaw}, which likely contradicts the mandates of many national research funding agencies.

This study represents a first step towards understanding group-level contributions to training data. Policymakers will likely want to consider additional analyses that extend this line of thinking. This could involve adapting similar methods to those we employed here, perhaps augmented with recent work on detecting pre-training data \cite{shiDetectingPretrainingData2023} or incorporating new methods for estimating racial disparities using surnames \cite{mccartanEstimatingRacialDisparities2023}, though this may require support from dataset creators and curators.

\subsection{Implications for Developers and Researchers of AI Systems}
At the highest level, the results above add to the growing list of reasons AI developers and researchers should consider shifting resources and attention away from the current LLM paradigm and towards both 1) LLM systems and techniques that only train on content with consent, and 2) mitigating the negative impacts that previous decision-making in the LLM industry may have caused. This paper highlights that those simply seeking to advance the current paradigm must reckon with the new structural biases against minority groups to which they may be contributing. They must also know that they are asking the Jewish community (and likely other minority communities as well) to make disproportionate sacrifices for the benefit of their mission. This is especially true for organizations like OpenAI, whose non-profit mandate requires that they build AI for the public good. More generally, like a number of other recent papers and opinion pieces \cite{orlowskiInternetOriginalSin2023,atwoodMurderedMyReplica2023}, this paper highlights that if we---as design researchers and practitioners---do not make significant changes to our approach, even if the community is successful at building something like an artificial general intelligence, such an accomplishment risks being forever tarnished with legitimacy issues originating from how it learned what it knows.

\subsection{Implications for Jewish Americans Who Author Digital Content}
Our results suggest a few strategies for action that can help Jewish Americans. It seems that if the groundwork were laid for easily accessible ``data opt out actions''---via national laws (e.g., \cite{knibbs_congress_2024}) or normatively adopted data use policies (e.g., \cite{miller_aitxt_2023})---Jewish Americans may be able to organize a very impactful opt out campaign, i.e., a ``data strike'' \cite{vincent_data_2019}. Additionally, this would suggest a natural incentive for AI firms to tackle the concerns laid out in this paper head on: if legal or technical tools for exerting data agency at the group level proliferate, any groups that currently see high levels of exposure to property dispossession could create significant leverage if there is sufficient buy-in (which may require pressuring or convincing institutions that own the rights to some members' content). These dynamics also mean that affinity and interest groups seeking to protect the welfare of minority groups---e.g., organizations that support the Jewish community---may have a natural alliance with efforts to promote a content generator-friendly AI paradigm.

\subsection{Implications for Other Minority Groups}
Our results suggest that Jewish Americans are not the only minority group that will likely serve as a ``canary in the coal mine'' and experience negative effects from the current LLM paradigm sooner than the general population. Much of the highest-value content used by language models requires significant education to create (e.g., consider the typical author of a paper in PubMed). Jewish Americans have a high relative average educational attainment \cite{murphyMostLeastEducated} and, although other factors may be involved (e.g., a tendency to choose careers involving more public knowledge sharing), that is likely one reason we saw the effects that we did. Jewish Americans, however, are of course far from the only minority group in the United States (let alone outside of it) with high average educational attainment: this is true of Asian Americans \cite{deptOfEdEducationalAttainment} and Hindu Americans \cite{murphyMostLeastEducated} as well, for instance. Assuming the link between education and valuable technical content for LLM training datasets is quite strong, Asian Americans and Hindu Americans are likely to be similarly affected by language models. Replicating and extending this work to examine the effect of non-consensual content training on these groups is a critical area of future work.

Our hypotheses are less clear regarding the potential effects of LLM-caused intellectual property dispossession for minority groups outside the United States. Countries around the world are rapidly engaging in policy discussions around language models, and doing work to understand if trends similar to those observed here affect minority groups in places like the EU, the UK, and elsewhere is also critical research that should happen quickly.

\subsection{Balancing Dispossession Concerns and Group-level Performance Gap Concerns}
There is a large body of early and field-defining work in algorithmic fairness that has highlighted issues with performance gaps that arise when minoritized groups are underrepresented in training data (see, e.g., Mehrabi et al.\ for a survey \cite{mehrabi2021survey}). Though concerns with under-representation could be seen as in tension with the argument we've put forth here, these two ideas are not mutually incompatible.

Generally, under-representation is most concerning when technologies downstream from the dataset impact people subject to the technologies. For instance, facial recognition systems---which can be used to unlock a phone screen or for policing---have been shown to have serious issues with respect to skin tone and gender~\cite{buolamwini2018gender}.

In the context of LLM-based technologies enabling labor substitution, however, data contributors do not necessarily derive utility from the technology nor are they subject to model outputs; rather, they are primarily subject to labor market dynamics.

This suggests that while there may be cases in which under-representation concerns dominate (and vice versa), in general the ML community will need to adopt a balanced approach to data representation. It can be simultaneously the case that some groups see harms related to lack of representation and others see harms because they have provided a large share of creative works. Ultimately, dataset curators are charged with the challenging but necessary task of gathering data that both accurately represents the world itself and accounts for the preferences of data subjects (and in some cases, includes some element of fair remuneration).

\subsection{Implications for HCI}
HCI has long been a leader in identifying and working to address challenges in the power dynamics between content producers and AI systems that consume their content. While HCI's contributions have primarily focused on crowd markets and their use in older generations of AI (e.g., \cite{alkhatib2017examining, kittur_future_2013, hara_data-driven_2018}), the research in this paper highlights the importance of continuing and strengthening this line of work in the LLM era. What kind of platforms could be created to help content producers receive a fair share of the value they are creating? (e.g., \cite{lampinen2017market, lampinen2018power}) How can we empower collective action among content producers to maximize their ability to do so? (e.g., \cite{li_out_2018}) These are all questions that can help to create new and more equitable LLM paradigms, and they require significant leadership from HCI researchers.

The large area of research addressing ``algorithmic bias'' can trace some of its origins to work at HCI venues like CHI (e.g., \cite{hecht_tower_2010}). This paper highlights the need for the HCI community to continue to push this line of work forward. While algorithmic bias research has mostly focused on ``representation harms'' \cite{crawford_trouble_2017} and gender and racial dimensions, this work highlights the urgent need to expand (but certainly not shift) our lens to consider direct material harms \cite{hecht_hci_2017, hecht_origins_2017} and additional dimensions.

HCI researchers have also looked at how we can best communicate AI design and fairness concepts to end users, including how the presentation of information about transparency can impact perceptions~\cite{vanberkel2021presentation}. For instance, Anik et al.~\cite{anik2021explaining} investigate how explanations of training data in ML systems can increase transparency and influence trustworthiness of systems. Further work in HCI could explore how disparate impacts and IP dispossession interact with perceptions of accuracy, trustworthiness, and fairness of LLMs. Similarly, HCI has also contributed to toolkits for ML practitioners to support better decision-making around fairness, bias, and transparency~\cite{lee2021landscape}. As one example, Madaio et al.~\cite{madaio2020codesign} co-designed AI fairness checklists with practitioners to ground them in the realities of the day-to-day work, paying particular attention to the organizational and sociotechnical factors that inhibit fairness work. Future work from HCI researchers might consider how dispossession and disproportionate impact on minority communities can be reckoned with within organizations.

Finally, as noted above, many of the harms from property dispossession identified in this paper can be mitigated if the net outcome of LLM-based technologies is to augment human labor rather than to substitute for it, a net outcome that the HCI community can help manifest. Indeed, if we were to be successful in doing so, groups may \textit{want} their data to appear in training sets as it could an enhance LLM's ability to augment their labor. While substituting for human labor is a core tenant of important LLM actors like OpenAI \cite{openai_openai_2018}, as discussed above, economists have suggested that large-scale net substitution is not a certainty (e.g., \cite{brynjolfssonTuringTrap}). These economists argue that researchers and practitioners in the computing community should focus on building LLM-based applications that keep human labor as input and create entirely new capabilities or augment existing ones, rather than building tools that implicitly seek to remove human inputs. In broad productivity terms, this means focusing on creating novel outputs rather than trying to drive human inputs to zero. As key problem-definers and developers of novel applications, as well as being a research community responsible for understanding users and their needs, the HCI community is well-positioned to lead the push for more augmentative LLM-based applications.
\section{Conclusion}
We find evidence that Jewish Americans likely have experienced a disproportionate share of the potential IP dispossession stemming from longstanding model training practices of AI companies and the broader AI research community. These practices may lead to serious economic harms with disturbing historical parallels, and call for urgent reflection about the future of the AI ecosystem. We discuss implications for a range of impacted groups and society as a whole.

\begin{acks}
The authors would like to thank the many other scholars---in the Jewish community and beyond---who took the time to give us feedback on this work. We in particular thank Glen Weyl, Moshe Vardi, and Gideon Taylor for their essential advice \& feedback on earlier drafts. We also thank Pearl Beck, Elizabeth Tighe, and Ira Sheskin for their invaluable inputs on demographic methods and DJN use. This research was not supported by institutions other than our home universities. Other lines of research by the authors have been supported by companies that produce LLMs and LLM-based applications including Microsoft, OpenAI, and Google.
\end{acks}

\balance
\bibliographystyle{ACM-Reference-Format}
\bibliography{main}


\begin{thebibliography}{133}


\ifx \showCODEN    \undefined \def \showCODEN     #1{\unskip}     \fi
\ifx \showDOI      \undefined \def \showDOI       #1{#1}\fi
\ifx \showISBNx    \undefined \def \showISBNx     #1{\unskip}     \fi
\ifx \showISBNxiii \undefined \def \showISBNxiii  #1{\unskip}     \fi
\ifx \showISSN     \undefined \def \showISSN      #1{\unskip}     \fi
\ifx \showLCCN     \undefined \def \showLCCN      #1{\unskip}     \fi
\ifx \shownote     \undefined \def \shownote      #1{#1}          \fi
\ifx \showarticletitle \undefined \def \showarticletitle #1{#1}   \fi
\ifx \showURL      \undefined \def \showURL       {\relax}        \fi
\providecommand\bibfield[2]{#2}
\providecommand\bibinfo[2]{#2}
\providecommand\natexlab[1]{#1}
\providecommand\showeprint[2][]{arXiv:#2}

\bibitem[Acemoglu et~al\mbox{.}(2022)]%
        {acemoglu2022artificial}
\bibfield{author}{\bibinfo{person}{Daron Acemoglu}, \bibinfo{person}{David Autor}, \bibinfo{person}{Jonathon Hazell}, {and} \bibinfo{person}{Pascual Restrepo}.} \bibinfo{year}{2022}\natexlab{}.
\newblock \showarticletitle{Artificial intelligence and jobs: evidence from online vacancies}.
\newblock \bibinfo{journal}{\emph{Journal of Labor Economics}} \bibinfo{volume}{40}, \bibinfo{number}{S1} (\bibinfo{year}{2022}), \bibinfo{pages}{S293--S340}.
\newblock


\bibitem[Achakulvisut et~al\mbox{.}(2020)]%
        {pubmed-parser}
\bibfield{author}{\bibinfo{person}{Titipat Achakulvisut}, \bibinfo{person}{Daniel~E. Acuna}, {and} \bibinfo{person}{Konrad Kording}.} \bibinfo{year}{2020}\natexlab{}.
\newblock \bibinfo{title}{Pubmed Parser: A Python Parser for PubMed Open-Access XML Subset and MEDLINE XML Dataset XML Dataset}.
\newblock
\newblock
\urldef\tempurl%
\url{https://doi.org/10.21105/joss.01979}
\showDOI{\tempurl}


\bibitem[Alkhatib et~al\mbox{.}(2017)]%
        {alkhatib2017examining}
\bibfield{author}{\bibinfo{person}{Ali Alkhatib}, \bibinfo{person}{Michael~S. Bernstein}, {and} \bibinfo{person}{Margaret Levi}.} \bibinfo{year}{2017}\natexlab{}.
\newblock \showarticletitle{Examining Crowd Work and Gig Work Through The Historical Lens of Piecework}. In \bibinfo{booktitle}{\emph{Proceedings of the 2017 CHI Conference on Human Factors in Computing Systems}} (Denver, Colorado, USA) \emph{(\bibinfo{series}{CHI '17})}. \bibinfo{publisher}{Association for Computing Machinery}, \bibinfo{address}{New York, NY, USA}, \bibinfo{pages}{4599–4616}.
\newblock
\showISBNx{9781450346559}
\urldef\tempurl%
\url{https://doi.org/10.1145/3025453.3025974}
\showDOI{\tempurl}


\bibitem[Altman(2021)]%
        {altmanMooresLaw}
\bibfield{author}{\bibinfo{person}{Sam Altman}.} \bibinfo{year}{2021}\natexlab{}.
\newblock \bibinfo{title}{Moore's Law for Everything}.
\newblock
\newblock
\urldef\tempurl%
\url{https://moores.samaltman.com/}
\showURL{%
\tempurl}


\bibitem[Andreadis(2022)]%
        {andreadisSeizureOfJewishIP}
\bibfield{author}{\bibinfo{person}{Steve Andreadis}.} \bibinfo{year}{2022}\natexlab{}.
\newblock \bibinfo{title}{The Seizure of Jewish Intellectual Property Ahead of World War II}.
\newblock
\newblock
\urldef\tempurl%
\url{https://blogs.loc.gov/copyright/2022/04/the-seizure-of-jewish-intellectual-property-ahead-of-world-war-ii}
\showURL{%
\tempurl}


\bibitem[Anik and Bunt(2021)]%
        {anik2021explaining}
\bibfield{author}{\bibinfo{person}{Ariful~Islam Anik} {and} \bibinfo{person}{Andrea Bunt}.} \bibinfo{year}{2021}\natexlab{}.
\newblock \showarticletitle{Data-Centric Explanations: Explaining Training Data of Machine Learning Systems to Promote Transparency}. In \bibinfo{booktitle}{\emph{Proceedings of the 2021 CHI Conference on Human Factors in Computing Systems}} \emph{(\bibinfo{series}{CHI '21})}. \bibinfo{publisher}{Association for Computing Machinery}, \bibinfo{address}{New York, NY, USA}, Article \bibinfo{articleno}{75}, \bibinfo{numpages}{13}~pages.
\newblock
\showISBNx{9781450380966}
\urldef\tempurl%
\url{https://doi.org/10.1145/3411764.3445736}
\showDOI{\tempurl}


\bibitem[Aronson et~al\mbox{.}(2016)]%
        {boxerAllPoliticsIsLocal2016}
\bibfield{author}{\bibinfo{person}{Janet~Krasner Aronson}, \bibinfo{person}{Matthew Boxer}, {and} \bibinfo{person}{Leonard Saxe}.} \bibinfo{year}{2016}\natexlab{}.
\newblock \showarticletitle{‘All Politics is Local’: Challenges in the Study of Local Jewish Communities}.
\newblock \bibinfo{journal}{\emph{Contemporary Jewry}} \bibinfo{volume}{36}, \bibinfo{number}{3} (\bibinfo{date}{Oct} \bibinfo{year}{2016}), \bibinfo{pages}{361–380}.
\newblock
\showISSN{1876-5165}
\urldef\tempurl%
\url{https://doi.org/10.1007/s12397-016-9200-7}
\showDOI{\tempurl}


\bibitem[Atwood(2023)]%
        {atwoodMurderedMyReplica2023}
\bibfield{author}{\bibinfo{person}{Margaret Atwood}.} \bibinfo{year}{2023}\natexlab{}.
\newblock \bibinfo{title}{Murdered by My Replica?}
\newblock
\newblock
\urldef\tempurl%
\url{https://www.theatlantic.com/books/archive/2023/08/ai-chatbot-training-books-margaret-atwood/675151/}
\showURL{%
\tempurl}


\bibitem[Bandy and Vincent(2021)]%
        {bandy2021addressing}
\bibfield{author}{\bibinfo{person}{John Bandy} {and} \bibinfo{person}{Nicholas Vincent}.} \bibinfo{year}{2021}\natexlab{}.
\newblock \showarticletitle{Addressing "Documentation Debt" in Machine Learning: A Retrospective Datasheet for BookCorpus}. In \bibinfo{booktitle}{\emph{Proceedings of the Neural Information Processing Systems Track on Datasets and Benchmarks}}, \bibfield{editor}{\bibinfo{person}{J.~Vanschoren} {and} \bibinfo{person}{S.~Yeung}} (Eds.), Vol.~\bibinfo{volume}{1}. \bibinfo{publisher}{Curran}, \bibinfo{address}{San Diego, CA, USA}.
\newblock
\urldef\tempurl%
\url{https://datasets-benchmarks-proceedings.neurips.cc/paper_files/paper/2021/file/54229abfcfa5649e7003b83dd4755294-Paper-round1.pdf}
\showURL{%
\tempurl}


\bibitem[Bannister(2022)]%
        {bannisterLetterJewishPropertyInSpain}
\bibfield{author}{\bibinfo{person}{Miranda Bannister}.} \bibinfo{year}{2022}\natexlab{}.
\newblock \bibinfo{title}{A 1492 Letter Regarding Jewish Property in Spain}.
\newblock
\newblock
\urldef\tempurl%
\url{https://mjhnyc.org/blog/1492-letter-regarding-jewish-property-in-spain/}
\showURL{%
\tempurl}


\bibitem[Barner(2017)]%
        {barnerAryanizationExpanded}
\bibfield{author}{\bibinfo{person}{Lida Barner}.} \bibinfo{year}{2017}\natexlab{}.
\newblock \bibinfo{booktitle}{\emph{“Aryanization” Expanded?: Patent Rights of Jews under the Nazi Regime}}.
\newblock \bibinfo{publisher}{Central European University Press}, \bibinfo{address}{Budapest, Hungary}, \bibinfo{pages}{127–144}.
\newblock
\showISBNx{978-963-386-185-1}
\urldef\tempurl%
\url{https://www.jstor.org/stable/10.7829/j.ctt1t6p66t.10}
\showURL{%
\tempurl}


\bibitem[Bender et~al\mbox{.}(2021)]%
        {bender2021}
\bibfield{author}{\bibinfo{person}{Emily~M. Bender}, \bibinfo{person}{Timnit Gebru}, \bibinfo{person}{Angelina McMillan-Major}, {and} \bibinfo{person}{Shmargaret Shmitchell}.} \bibinfo{year}{2021}\natexlab{}.
\newblock \showarticletitle{On the Dangers of Stochastic Parrots: Can Language Models Be Too Big?}. In \bibinfo{booktitle}{\emph{Proceedings of the 2021 ACM Conference on Fairness, Accountability, and Transparency}} (Virtual Event, Canada) \emph{(\bibinfo{series}{FAccT '21})}. \bibinfo{publisher}{Association for Computing Machinery}, \bibinfo{address}{New York, NY, USA}, \bibinfo{pages}{610–623}.
\newblock
\showISBNx{9781450383097}
\urldef\tempurl%
\url{https://doi.org/10.1145/3442188.3445922}
\showDOI{\tempurl}


\bibitem[Black(2012)]%
        {black_ibm_2012}
\bibfield{author}{\bibinfo{person}{Edwin Black}.} \bibinfo{year}{2012}\natexlab{}.
\newblock \bibinfo{booktitle}{\emph{{IBM} and the Holocaust: The Strategic Alliance Between Nazi Germany and America's Most Powerful Corporation-Expanded Edition} (\bibinfo{edition}{2nd edition} ed.)}.
\newblock \bibinfo{publisher}{Dialog Press}, \bibinfo{address}{USA}.
\newblock
\showISBNx{978-0-914153-27-6}


\bibitem[Black et~al\mbox{.}(2022)]%
        {blackGPTNeoX20BOpenSourceAutoregressive2022}
\bibfield{author}{\bibinfo{person}{Sid Black}, \bibinfo{person}{Stella Biderman}, \bibinfo{person}{Eric Hallahan}, \bibinfo{person}{Quentin Anthony}, \bibinfo{person}{Leo Gao}, \bibinfo{person}{Laurence Golding}, \bibinfo{person}{Horace He}, \bibinfo{person}{Connor Leahy}, \bibinfo{person}{Kyle McDonell}, \bibinfo{person}{Jason Phang}, \bibinfo{person}{Michael Pieler}, \bibinfo{person}{USVSN~Sai Prashanth}, \bibinfo{person}{Shivanshu Purohit}, \bibinfo{person}{Laria Reynolds}, \bibinfo{person}{Jonathan Tow}, \bibinfo{person}{Ben Wang}, {and} \bibinfo{person}{Samuel Weinbach}.} \bibinfo{year}{2022}\natexlab{}.
\newblock \bibinfo{title}{{GPT}-{NeoX}-{20B}: {An} {Open}-{Source} {Autoregressive} {Language} {Model}}.
\newblock
\newblock
\urldef\tempurl%
\url{https://doi.org/10.48550/arXiv.2204.06745}
\showDOI{\tempurl}
\newblock
\shownote{arXiv:2204.06745 [cs]}.


\bibitem[Bogle(2023)]%
        {bogleNewsBlockOAICrawler}
\bibfield{author}{\bibinfo{person}{Ariel Bogle}.} \bibinfo{year}{2023}\natexlab{}.
\newblock \bibinfo{title}{New York Times, CNN and Australia’s ABC block OpenAI’s GPTBot web crawler from accessing content}.
\newblock
\newblock
\showISSN{0261-3077}
\urldef\tempurl%
\url{https://www.theguardian.com/technology/2023/aug/25/new-york-times-cnn-and-abc-block-openais-gptbot-web-crawler-from-scraping-content}
\showURL{%
\tempurl}


\bibitem[Brynjolfsson(2022)]%
        {brynjolfssonTuringTrap}
\bibfield{author}{\bibinfo{person}{Erik Brynjolfsson}.} \bibinfo{year}{2022}\natexlab{}.
\newblock \bibinfo{title}{The Turing Trap: The Promise and Peril of Human-Like Artificial Intelligence}.
\newblock
\newblock
\urldef\tempurl%
\url{https://digitaleconomy.stanford.edu/news/the-turing-trap-the-promise-peril-of-human-like-artificial-intelligence/}
\showURL{%
\tempurl}


\bibitem[Buolamwini and Gebru(2018)]%
        {buolamwini2018gender}
\bibfield{author}{\bibinfo{person}{Joy Buolamwini} {and} \bibinfo{person}{Timnit Gebru}.} \bibinfo{year}{2018}\natexlab{}.
\newblock \showarticletitle{Gender Shades: Intersectional Accuracy Disparities in Commercial Gender Classification}. In \bibinfo{booktitle}{\emph{Proceedings of the 1st Conference on Fairness, Accountability and Transparency}} \emph{(\bibinfo{series}{Proceedings of Machine Learning Research}, Vol.~\bibinfo{volume}{81})}, \bibfield{editor}{\bibinfo{person}{Sorelle~A. Friedler} {and} \bibinfo{person}{Christo Wilson}} (Eds.). \bibinfo{publisher}{PMLR}, \bibinfo{address}{New York, NY, USA}, \bibinfo{pages}{77--91}.
\newblock
\urldef\tempurl%
\url{https://proceedings.mlr.press/v81/buolamwini18a.html}
\showURL{%
\tempurl}


\bibitem[Carlos et~al\mbox{.}(2022)]%
        {carlosIndigenousNationsDevelopment2022}
\bibfield{author}{\bibinfo{person}{Ann~M. Carlos}, \bibinfo{person}{Donna~L. Feir}, {and} \bibinfo{person}{Angela Redish}.} \bibinfo{year}{2022}\natexlab{}.
\newblock \showarticletitle{Indigenous {Nations} and the {Development} of the {U}.{S}. {Economy}: {Land}, {Resources}, and {Dispossession}}.
\newblock \bibinfo{journal}{\emph{The Journal of Economic History}} \bibinfo{volume}{82}, \bibinfo{number}{2} (\bibinfo{date}{June} \bibinfo{year}{2022}), \bibinfo{pages}{516--555}.
\newblock
\showISSN{0022-0507, 1471-6372}
\urldef\tempurl%
\url{https://doi.org/10.1017/S0022050722000080}
\showDOI{\tempurl}
\newblock
\shownote{Publisher: Cambridge University Press}.


\bibitem[Castro~Fernandez(2023)]%
        {castrofernandezDataSharingMarketsModel2023}
\bibfield{author}{\bibinfo{person}{Raul Castro~Fernandez}.} \bibinfo{year}{2023}\natexlab{}.
\newblock \showarticletitle{Data-{Sharing} {Markets}: {Model}, {Protocol}, and {Algorithms} to {Incentivize} the {Formation} of {Data}-{Sharing} {Consortia}}.
\newblock \bibinfo{journal}{\emph{Proceedings of the ACM on Management of Data}} \bibinfo{volume}{1}, \bibinfo{number}{2} (\bibinfo{date}{June} \bibinfo{year}{2023}), \bibinfo{pages}{172:1--172:25}.
\newblock
\urldef\tempurl%
\url{https://doi.org/10.1145/3589317}
\showDOI{\tempurl}


\bibitem[Census.gov(2021)]%
        {us-census-list}
\bibfield{author}{\bibinfo{person}{Census.gov}.} \bibinfo{year}{2021}\natexlab{}.
\newblock \bibinfo{title}{Frequently Occurring Surnames from the 2010 Census}.
\newblock
\newblock
\urldef\tempurl%
\url{https://www.census.gov/topics/population/genealogy/data/2010_surnames.html}
\showURL{%
\tempurl}


\bibitem[Center(2013)]%
        {pew2013Report}
\bibfield{author}{\bibinfo{person}{Pew~Research Center}.} \bibinfo{year}{2013}\natexlab{}.
\newblock \bibinfo{title}{A portrait of Jewish Americans: Findings from a Pew research center survey of US Jews}.
\newblock
\newblock


\bibitem[Center(2021)]%
        {pew2020Report}
\bibfield{author}{\bibinfo{person}{Pew~Research Center}.} \bibinfo{year}{2021}\natexlab{}.
\newblock \bibinfo{title}{Jewish Americans in 2020}.
\newblock
\newblock
\urldef\tempurl%
\url{https://www.pewresearch.org/religion/2021/05/11/jewish-americans-in-2020/}
\showURL{%
\tempurl}


\bibitem[Central(2023)]%
        {pmcData}
\bibfield{author}{\bibinfo{person}{PubMed Central}.} \bibinfo{year}{2023}\natexlab{}.
\newblock \bibinfo{title}{PMC Open Access Subset}.
\newblock
\newblock
\urldef\tempurl%
\url{https://www.ncbi.nlm.nih.gov/pmc/tools/openftlist/}
\showURL{%
\tempurl}


\bibitem[Clement et~al\mbox{.}(2019a)]%
        {arxiv-public-datasets}
\bibfield{author}{\bibinfo{person}{Colin~B. Clement}, \bibinfo{person}{Matthew Bierbaum}, \bibinfo{person}{Kevin~P. O'Keeffe}, {and} \bibinfo{person}{Alexander~A. Alemi}.} \bibinfo{year}{2019}\natexlab{a}.
\newblock \bibinfo{title}{arxiv-public-datasets}.
\newblock
\newblock
\urldef\tempurl%
\url{https://github.com/mattbierbaum/arxiv-public-datasets/blob/master/README.md}
\showURL{%
\tempurl}


\bibitem[Clement et~al\mbox{.}(2019b)]%
        {arxiv-public-datasets-Paper}
\bibfield{author}{\bibinfo{person}{Colin~B. Clement}, \bibinfo{person}{Matthew Bierbaum}, \bibinfo{person}{Kevin~P. O’Keeffe}, {and} \bibinfo{person}{Alexander~A. Alemi}.} \bibinfo{year}{2019}\natexlab{b}.
\newblock \showarticletitle{On the Use of ArXiv as a Dataset}.
\newblock \bibinfo{journal}{\emph{ArXiv}}  \bibinfo{volume}{abs/1905.00075} (\bibinfo{year}{2019}), \bibinfo{pages}{1--7}.
\newblock
\urldef\tempurl%
\url{https://api.semanticscholar.org/CorpusID:141496572}
\showURL{%
\tempurl}


\bibitem[Cohen(2016)]%
        {cohenDeficientIfNotDistorted2016}
\bibfield{author}{\bibinfo{person}{Steven~M. Cohen}.} \bibinfo{year}{2016}\natexlab{}.
\newblock \showarticletitle{Deficient, If Not Distorted: Jewish Community Studies That Totally Rely upon Known Jewish Households}.
\newblock \bibinfo{journal}{\emph{Contemporary Jewry}} \bibinfo{volume}{36}, \bibinfo{number}{3} (\bibinfo{date}{Oct} \bibinfo{year}{2016}), \bibinfo{pages}{343–360}.
\newblock
\showISSN{1876-5165}
\urldef\tempurl%
\url{https://doi.org/10.1007/s12397-016-9187-0}
\showDOI{\tempurl}


\bibitem[Cohen et~al\mbox{.}(2001)]%
        {njps2000Data}
\bibfield{author}{\bibinfo{person}{Steven~M. Cohen}, \bibinfo{person}{Frank Mott}, \bibinfo{person}{Lorraine Blass}, \bibinfo{person}{Jim Schwartz}, \bibinfo{person}{Jonathon Ament}, \bibinfo{person}{Vivian Klaff}, {and} \bibinfo{person}{Laurence Kotler-Berkowitz}.} \bibinfo{year}{2001}\natexlab{}.
\newblock \bibinfo{title}{2000-01 National Jewish Population Survey}.
\newblock
\newblock
\urldef\tempurl%
\url{https://www.jewishdatabank.org/databank/search-results/study/307}
\showURL{%
\tempurl}


\bibitem[CourtListener(2010)]%
        {courtListener-API}
\bibfield{author}{\bibinfo{person}{CourtListener}.} \bibinfo{year}{2010}\natexlab{}.
\newblock \bibinfo{title}{CourtListener}.
\newblock
\newblock
\urldef\tempurl%
\url{https://www.courtlistener.com/help/api/bulk-data/}
\showURL{%
\tempurl}


\bibitem[Craemer et~al\mbox{.}(2020)]%
        {craemerWealthImplicationsSlavery2020}
\bibfield{author}{\bibinfo{person}{Thomas Craemer}, \bibinfo{person}{Trevor Smith}, \bibinfo{person}{Brianna Harrison}, \bibinfo{person}{Trevon Logan}, \bibinfo{person}{Wesley Bellamy}, {and} \bibinfo{person}{William Darity}.} \bibinfo{year}{2020}\natexlab{}.
\newblock \showarticletitle{Wealth {Implications} of {Slavery} and {Racial} {Discrimination} for {African} {American} {Descendants} of the {Enslaved}}.
\newblock \bibinfo{journal}{\emph{The Review of Black Political Economy}} \bibinfo{volume}{47}, \bibinfo{number}{3} (\bibinfo{date}{Sept.} \bibinfo{year}{2020}), \bibinfo{pages}{218--254}.
\newblock
\showISSN{0034-6446}
\urldef\tempurl%
\url{https://doi.org/10.1177/0034644620926516}
\showDOI{\tempurl}
\newblock
\shownote{Publisher: SAGE Publications Inc}.


\bibitem[Cramer(2021)]%
        {cramerHenriettaLacks}
\bibfield{author}{\bibinfo{person}{Maria Cramer}.} \bibinfo{year}{2021}\natexlab{}.
\newblock \bibinfo{title}{Henrietta Lacks, Whose Cells Were Taken Without Her Consent, Is Honored by W.H.O.}
\newblock
\newblock
\showISSN{0362-4331}
\urldef\tempurl%
\url{https://www.nytimes.com/2021/10/13/science/henrietta-lacks-cells-who.html}
\showURL{%
\tempurl}


\bibitem[Crawford(2017)]%
        {crawford_trouble_2017}
\bibfield{author}{\bibinfo{person}{Kate Crawford}.} \bibinfo{year}{2017}\natexlab{}.
\newblock \bibinfo{booktitle}{\emph{The {Trouble} {With} {Bias}}}.
\newblock The Annual Conference on Neural Information Processing Systems (NeruIPS), San Diego, CA, USA.
\newblock


\bibitem[Davis(2023)]%
        {davisSarahSilvermanSuing2023}
\bibfield{author}{\bibinfo{person}{Wes Davis}.} \bibinfo{year}{2023}\natexlab{}.
\newblock \bibinfo{title}{Sarah {Silverman} is suing {OpenAI} and {Meta} for copyright infringement}.
\newblock
\newblock
\urldef\tempurl%
\url{https://www.theverge.com/2023/7/9/23788741/sarah-silverman-openai-meta-chatgpt-llama-copyright-infringement-chatbots-artificial-intelligence-ai}
\showURL{%
\tempurl}


\bibitem[DellaPergola(2022)]%
        {dellaPergolaAJYB2021}
\bibfield{author}{\bibinfo{person}{Sergio DellaPergola}.} \bibinfo{year}{2022}\natexlab{}.
\newblock \bibinfo{booktitle}{\emph{World Jewish Population, 2021}}.
\newblock \bibinfo{publisher}{Springer International Publishing}, \bibinfo{address}{Cham}, \bibinfo{pages}{313–412}.
\newblock
\showISBNx{978-3-030-99750-2}
\urldef\tempurl%
\url{https://doi.org/10.1007/978-3-030-99750-2_8}
\showDOI{\tempurl}


\bibitem[Diakopoulos(2023)]%
        {diakopoulosFindingEvidenceMemorized2023}
\bibfield{author}{\bibinfo{person}{Nick Diakopoulos}.} \bibinfo{year}{2023}\natexlab{}.
\newblock \bibinfo{title}{Finding {Evidence} of {Memorized} {News} {Content} in {GPT} {Models}}.
\newblock
\newblock
\urldef\tempurl%
\url{https://generative-ai-newsroom.com/finding-evidence-of-memorized-news-content-in-gpt-models-d11a73576d2}
\showURL{%
\tempurl}


\bibitem[D'Ignazio and Klein(2020)]%
        {dignazio2020what}
\bibfield{author}{\bibinfo{person}{Catherine D'Ignazio} {and} \bibinfo{person}{Lauren Klein}.} \bibinfo{year}{2020}\natexlab{}.
\newblock \bibinfo{booktitle}{\emph{4. ``{What} {Gets} {Counted} {Counts}''}}.
\newblock \bibinfo{publisher}{MIT Press}, \bibinfo{address}{Cambridge, MA, USA}, Chapter~4, \bibinfo{pages}{1--27}.
\newblock
\urldef\tempurl%
\url{https://data-feminism.mitpress.mit.edu/pub/h1w0nbqp}
\showURL{%
\tempurl}


\bibitem[Dutwin(2016)]%
        {dutwinEverythingToConsider2016}
\bibfield{author}{\bibinfo{person}{David Dutwin}.} \bibinfo{year}{2016}\natexlab{}.
\newblock \showarticletitle{Everything You Need to Consider When Deciding to Field a Survey of Jews: Choices in Survey Methods and Their Consequences on Quality}.
\newblock \bibinfo{journal}{\emph{Contemporary Jewry}} \bibinfo{volume}{36}, \bibinfo{number}{3} (\bibinfo{date}{Oct} \bibinfo{year}{2016}), \bibinfo{pages}{297–318}.
\newblock
\showISSN{1876-5165}
\urldef\tempurl%
\url{https://doi.org/10.1007/s12397-016-9189-y}
\showDOI{\tempurl}


\bibitem[EleutherAI(2020)]%
        {github-downloader}
\bibfield{author}{\bibinfo{person}{EleutherAI}.} \bibinfo{year}{2020}\natexlab{}.
\newblock \bibinfo{title}{github-downloader}.
\newblock
\newblock
\urldef\tempurl%
\url{https://github.com/EleutherAI/github-downloader}
\showURL{%
\tempurl}


\bibitem[EleutherAI(2023)]%
        {eleutherAIHomepage}
\bibfield{author}{\bibinfo{person}{EleutherAI}.} \bibinfo{year}{2023}\natexlab{}.
\newblock \bibinfo{title}{EleutherAI}.
\newblock
\newblock
\urldef\tempurl%
\url{https://www.eleuther.ai}
\showURL{%
\tempurl}


\bibitem[Eloundou et~al\mbox{.}(2023)]%
        {eloundou_gpts_2023}
\bibfield{author}{\bibinfo{person}{Tyna Eloundou}, \bibinfo{person}{Sam Manning}, \bibinfo{person}{Pamela Mishkin}, {and} \bibinfo{person}{Daniel Rock}.} \bibinfo{year}{2023}\natexlab{}.
\newblock \bibinfo{title}{{GPTs} are {GPTs}: An Early Look at the Labor Market Impact Potential of Large Language Models}.
\newblock
\newblock
\urldef\tempurl%
\url{https://doi.org/10.48550/arXiv.2303.10130}
\showDOI{\tempurl}
\showeprint[arxiv]{2303.10130 [cs, econ, q-fin]}


\bibitem[Fairlie(2009)]%
        {fairlieShortHistoryOfEnclosure}
\bibfield{author}{\bibinfo{person}{Simon Fairlie}.} \bibinfo{year}{2009}\natexlab{}.
\newblock \bibinfo{title}{A Short History of Enclosure in Britain}.
\newblock
\newblock
\urldef\tempurl%
\url{https://www.thelandmagazine.org.uk/articles/short-history-enclosure-britain}
\showURL{%
\tempurl}


\bibitem[for Education~Statistics(2023)]%
        {deptOfEdEducationalAttainment}
\bibfield{author}{\bibinfo{person}{National~Center for Education~Statistics}.} \bibinfo{year}{2023}\natexlab{}.
\newblock \bibinfo{title}{Educational Attainment of Young Adults}.
\newblock
\newblock
\urldef\tempurl%
\url{https://nces.ed.gov/programs/coe/indicator/caa/young-adult-attainment}
\showURL{%
\tempurl}


\bibitem[Freedland(2018)]%
        {freedland_antisemitism_2018}
\bibfield{author}{\bibinfo{person}{Jonathan Freedland}.} \bibinfo{year}{2018}\natexlab{}.
\newblock \bibinfo{title}{Antisemitism matters: {Jews} are the canary in the coalmine}.
\newblock
\newblock
\showISSN{0261-3077}
\urldef\tempurl%
\url{https://www.theguardian.com/commentisfree/2018/mar/30/antisemitism-jews-canary-coalmine-fake-news}
\showURL{%
\tempurl}


\bibitem[Gao et~al\mbox{.}(2021)]%
        {thePile}
\bibfield{author}{\bibinfo{person}{Leo Gao}, \bibinfo{person}{Stella Biderman}, \bibinfo{person}{Sid Black}, \bibinfo{person}{Laurence Golding}, \bibinfo{person}{Travis Hoppe}, \bibinfo{person}{Charles Foster}, \bibinfo{person}{Jason Phang}, \bibinfo{person}{Horace He}, \bibinfo{person}{Anish Thite}, \bibinfo{person}{Noa Nabeshima}, \bibinfo{person}{Shawn Presser}, {and} \bibinfo{person}{Connor Leahy}.} \bibinfo{year}{2021}\natexlab{}.
\newblock \showarticletitle{The Pile: An 800GB Dataset of Diverse Text for Language Modeling}.
\newblock \bibinfo{journal}{\emph{CoRR}}  \bibinfo{volume}{abs/2101.00027} (\bibinfo{year}{2021}), \bibinfo{pages}{1--39}.
\newblock
\showeprint[arXiv]{2101.00027}
\urldef\tempurl%
\url{https://arxiv.org/abs/2101.00027}
\showURL{%
\tempurl}


\bibitem[Gebru et~al\mbox{.}(2021)]%
        {gebru2021datasheets}
\bibfield{author}{\bibinfo{person}{Timnit Gebru}, \bibinfo{person}{Jamie Morgenstern}, \bibinfo{person}{Briana Vecchione}, \bibinfo{person}{Jennifer~Wortman Vaughan}, \bibinfo{person}{Hanna Wallach}, \bibinfo{person}{Hal~Daum{\'e} III}, {and} \bibinfo{person}{Kate Crawford}.} \bibinfo{year}{2021}\natexlab{}.
\newblock \showarticletitle{Datasheets for Datasets}.
\newblock \bibinfo{journal}{\emph{Commun. ACM}} \bibinfo{volume}{64}, \bibinfo{number}{12} (\bibinfo{year}{2021}), \bibinfo{pages}{86--92}.
\newblock


\bibitem[GitHub(2022)]%
        {githubAPI}
\bibfield{author}{\bibinfo{person}{GitHub}.} \bibinfo{year}{2022}\natexlab{}.
\newblock \bibinfo{title}{GitHub REST API Documentation}.
\newblock
\newblock
\urldef\tempurl%
\url{https://docs.github.com/en/rest?apiVersion=2022-11-28}
\showURL{%
\tempurl}


\bibitem[Goldsmith(2023)]%
        {goldsmithMichaelChabonDavid2023}
\bibfield{author}{\bibinfo{person}{Jill Goldsmith}.} \bibinfo{year}{2023}\natexlab{}.
\newblock \bibinfo{title}{Michael Chabon, David Henry Hwang, Other Writers Sue Meta AI Platform LLaMA For Copyright Infringement, Seek Class Action Status}.
\newblock
\newblock
\urldef\tempurl%
\url{https://deadline.com/2023/09/michael-chabon-david-henry-hwang-writers-sue-meta-ai-llama-copyright-1235544842/}
\showURL{%
\tempurl}


\bibitem[Grynbaum and Mac(2023)]%
        {grynbaum_times_2023}
\bibfield{author}{\bibinfo{person}{Michael~M. Grynbaum} {and} \bibinfo{person}{Ryan Mac}.} \bibinfo{year}{2023}\natexlab{}.
\newblock \bibinfo{title}{The {Times} {Sues} {OpenAI} and {Microsoft} {Over} {A}.{I}. {Use} of {Copyrighted} {Work}}.
\newblock
\newblock
\showISSN{0362-4331}
\urldef\tempurl%
\url{https://www.nytimes.com/2023/12/27/business/media/new-york-times-open-ai-microsoft-lawsuit.html}
\showURL{%
\tempurl}


\bibitem[Guild(2023)]%
        {theauthorsguildAuthorsGuildSubmits2023}
\bibfield{author}{\bibinfo{person}{The~Authors Guild}.} \bibinfo{year}{2023}\natexlab{}.
\newblock \bibinfo{title}{Authors {Guild} {Submits} {Written} {Testimony} in {Senate} {AI} {Hearing}}.
\newblock
\newblock
\urldef\tempurl%
\url{https://authorsguild.org/news/ag-submits-written-testimony-in-senate-ai-hearing/}
\showURL{%
\tempurl}


\bibitem[Hamidi et~al\mbox{.}(2018)]%
        {hamidi2018gender}
\bibfield{author}{\bibinfo{person}{Foad Hamidi}, \bibinfo{person}{Morgan~Klaus Scheuerman}, {and} \bibinfo{person}{Stacy~M. Branham}.} \bibinfo{year}{2018}\natexlab{}.
\newblock \showarticletitle{Gender Recognition or Gender Reductionism? The Social Implications of Embedded Gender Recognition Systems}. In \bibinfo{booktitle}{\emph{Proceedings of the 2018 CHI Conference on Human Factors in Computing Systems}} \emph{(\bibinfo{series}{CHI ’18})}. \bibinfo{publisher}{Association for Computing Machinery}, \bibinfo{address}{New York, NY, USA}, \bibinfo{pages}{1–13}.
\newblock
\showISBNx{978-1-4503-5620-6}
\urldef\tempurl%
\url{https://doi.org/10.1145/3173574.3173582}
\showDOI{\tempurl}


\bibitem[Hara et~al\mbox{.}(2018)]%
        {hara_data-driven_2018}
\bibfield{author}{\bibinfo{person}{Kotaro Hara}, \bibinfo{person}{Abigail Adams}, \bibinfo{person}{Kristy Milland}, \bibinfo{person}{Saiph Savage}, \bibinfo{person}{Chris Callison-Burch}, {and} \bibinfo{person}{Jeffrey~P. Bigham}.} \bibinfo{year}{2018}\natexlab{}.
\newblock \showarticletitle{A {Data}-{Driven} {Analysis} of {Workers}' {Earnings} on {Amazon} {Mechanical} {Turk}}. In \bibinfo{booktitle}{\emph{Proceedings of the 2018 {CHI} {Conference} on {Human} {Factors} in {Computing} {Systems}}} \emph{(\bibinfo{series}{{CHI} '18})}. \bibinfo{publisher}{Association for Computing Machinery}, \bibinfo{address}{New York, NY, USA}, \bibinfo{pages}{1--14}.
\newblock
\showISBNx{978-1-4503-5620-6}
\urldef\tempurl%
\url{https://doi.org/10.1145/3173574.3174023}
\showDOI{\tempurl}


\bibitem[Hartman and Sheskin(2013)]%
        {sheskinCollegeCampus2013}
\bibfield{author}{\bibinfo{person}{Harriet Hartman} {and} \bibinfo{person}{Ira~M Sheskin}.} \bibinfo{year}{2013}\natexlab{}.
\newblock \showarticletitle{Estimating the Jewish student population of a college campus}.
\newblock \bibinfo{journal}{\emph{Journal of Jewish Communal Service}} \bibinfo{volume}{88}, \bibinfo{number}{1-2} (\bibinfo{year}{2013}), \bibinfo{pages}{95--109}.
\newblock


\bibitem[Hatzius et~al\mbox{.}(2023)]%
        {hatzius_potentially_2023}
\bibfield{author}{\bibinfo{person}{Jan Hatzius}, \bibinfo{person}{Joseph Briggs}, \bibinfo{person}{Devesh Kodnani}, {and} \bibinfo{person}{Giovanni Pierdomenico}.} \bibinfo{year}{2023}\natexlab{}.
\newblock \bibinfo{title}{The Potentially Large Effects of Artificial Intelligence on Economic Growth}.
\newblock
\newblock
\urldef\tempurl%
\url{https://www.gspublishing.com/content/research/en/reports/2023/03/27/d64e052b-0f6e-45d7-967b-d7be35fabd16.html}
\showURL{%
\tempurl}


\bibitem[Hecht(2017a)]%
        {hecht_hci_2017}
\bibfield{author}{\bibinfo{person}{Brent Hecht}.} \bibinfo{year}{2017}\natexlab{a}.
\newblock \showarticletitle{{HCI} and the {U}.{S}. {Presidential} {Election}: {A} {Few} {Thoughts} on a {Research} {Agenda}}. In \bibinfo{booktitle}{\emph{{CHI} '18 {Panel} {Presentation}: {The} 2016 {US} {Election} and {HCI}: {Towards} a {Research} {Agenda}}}. \bibinfo{publisher}{The Conference on Human Factors in Computing Systems (CHI)}, \bibinfo{address}{Denver, CO}, \bibinfo{pages}{1--5}.
\newblock
\urldef\tempurl%
\url{https://brenthecht.com/publications/chi17_bhecht_election2016panel.pdf}
\showURL{%
\tempurl}


\bibitem[Hecht(2017b)]%
        {hecht_origins_2017}
\bibfield{author}{\bibinfo{person}{Brent Hecht}.} \bibinfo{year}{2017}\natexlab{b}.
\newblock \bibinfo{title}{The Origins, Present, and Future of Algorithmic Bias}.  (\bibinfo{year}{2017}).
\newblock


\bibitem[Hecht and Gergle(2010)]%
        {hecht_tower_2010}
\bibfield{author}{\bibinfo{person}{Brent Hecht} {and} \bibinfo{person}{Darren Gergle}.} \bibinfo{year}{2010}\natexlab{}.
\newblock \showarticletitle{The {Tower} of {Babel} {Meets} {Web} 2.0: {User}-{Generated} {Content} and {Its} {Applications} in a {Multilingual} {Context}}. In \bibinfo{booktitle}{\emph{{CHI} '10: 28th {International} {Conference} on {Human} {Factors} in {Computing} {Systems}}} \emph{(\bibinfo{series}{{CHI} '10})}. \bibinfo{publisher}{ACM}, \bibinfo{address}{Atlanta, GA}, \bibinfo{pages}{291--300}.
\newblock
\showISBNx{978-1-60558-929-9}
\urldef\tempurl%
\url{https://doi.org/10.1145/1753326.1753370}
\showDOI{\tempurl}
\newblock
\shownote{ACM ID: 1753370}.


\bibitem[Himmelfarb(1986)]%
        {himmelfarbFurtherComments1876}
\bibfield{author}{\bibinfo{person}{Harold~S. Himmelfarb}.} \bibinfo{year}{1986}\natexlab{}.
\newblock \showarticletitle{Further comments on the use of DJN}.
\newblock \bibinfo{journal}{\emph{Contemporary Jewry}} \bibinfo{volume}{7}, \bibinfo{number}{1} (\bibinfo{date}{Jan} \bibinfo{year}{1986}), \bibinfo{pages}{99–102}.
\newblock
\showISSN{1876-5165}
\urldef\tempurl%
\url{https://doi.org/10.1007/BF02967946}
\showDOI{\tempurl}


\bibitem[Himmelfarb et~al\mbox{.}(1983)]%
        {himmelfarbSamplingByEthnicSurnames}
\bibfield{author}{\bibinfo{person}{Harold~S. Himmelfarb}, \bibinfo{person}{R.~Michael Loar}, {and} \bibinfo{person}{Susan~H. Mott}.} \bibinfo{year}{1983}\natexlab{}.
\newblock \showarticletitle{Sampling by Ethnic Surnames: The Case of American Jews}.
\newblock \bibinfo{journal}{\emph{The Public Opinion Quarterly}} \bibinfo{volume}{47}, \bibinfo{number}{2} (\bibinfo{year}{1983}), \bibinfo{pages}{247–260}.
\newblock
\showISSN{0033-362X}
\urldef\tempurl%
\url{http://www.jstor.org/stable/2749024}
\showURL{%
\tempurl}


\bibitem[Javed(2023)]%
        {javedMediaMogulWarns2023}
\bibfield{author}{\bibinfo{person}{Ayesha Javed}.} \bibinfo{year}{2023}\natexlab{}.
\newblock \bibinfo{title}{AI Could Destroy Journalism as We Know It. Media Mogul Barry Diller Hopes to Save It}.
\newblock
\newblock
\urldef\tempurl%
\url{https://time.com/6279147/barry-diller-ai-journalism/}
\showURL{%
\tempurl}


\bibitem[Jiang et~al\mbox{.}(2023)]%
        {jiangAIArtIts2023}
\bibfield{author}{\bibinfo{person}{Harry~H. Jiang}, \bibinfo{person}{Lauren Brown}, \bibinfo{person}{Jessica Cheng}, \bibinfo{person}{Mehtab Khan}, \bibinfo{person}{Abhishek Gupta}, \bibinfo{person}{Deja Workman}, \bibinfo{person}{Alex Hanna}, \bibinfo{person}{Johnathan Flowers}, {and} \bibinfo{person}{Timnit Gebru}.} \bibinfo{year}{2023}\natexlab{}.
\newblock \showarticletitle{{AI} {Art} and its {Impact} on {Artists}}. In \bibinfo{booktitle}{\emph{Proceedings of the 2023 {AAAI}/{ACM} {Conference} on {AI}, {Ethics}, and {Society}}} \emph{(\bibinfo{series}{{AIES} '23})}. \bibinfo{publisher}{Association for Computing Machinery}, \bibinfo{address}{New York, NY, USA}, \bibinfo{pages}{363--374}.
\newblock
\showISBNx{9798400702310}
\urldef\tempurl%
\url{https://doi.org/10.1145/3600211.3604681}
\showDOI{\tempurl}


\bibitem[{Kevin Roose} and {Casey Newton}(2023)]%
        {kevin_roose_casey_2023}
\bibfield{author}{\bibinfo{person}{{Kevin Roose}} {and} \bibinfo{person}{{Casey Newton}}.} \bibinfo{year}{2023}\natexlab{}.
\newblock \bibinfo{title}{Casey Goes to the White House + The Copyright Battle Over Artificial Intelligence + {HatGPT}}.
\newblock
\newblock
\urldef\tempurl%
\url{https://www.nytimes.com/2023/11/03/podcasts/hard-fork-executive-order-ai-copyright.html?}
\showURL{%
\tempurl}


\bibitem[Keyes(2018)]%
        {keyes2018misgendering}
\bibfield{author}{\bibinfo{person}{Os Keyes}.} \bibinfo{year}{2018}\natexlab{}.
\newblock \showarticletitle{The Misgendering Machines: Trans/HCI Implications of Automatic Gender Recognition}.
\newblock \bibinfo{journal}{\emph{Proceedings of the ACM on Human-Computer Interaction}} \bibinfo{volume}{2}, \bibinfo{number}{CSCW} (\bibinfo{date}{Nov} \bibinfo{year}{2018}), \bibinfo{pages}{88:1--88:22}.
\newblock
\urldef\tempurl%
\url{https://doi.org/10.1145/3274357}
\showDOI{\tempurl}


\bibitem[Kittur et~al\mbox{.}(2013)]%
        {kittur_future_2013}
\bibfield{author}{\bibinfo{person}{Aniket Kittur}, \bibinfo{person}{Jeffrey~V. Nickerson}, \bibinfo{person}{Michael Bernstein}, \bibinfo{person}{Elizabeth Gerber}, \bibinfo{person}{Aaron Shaw}, \bibinfo{person}{John Zimmerman}, \bibinfo{person}{Matt Lease}, {and} \bibinfo{person}{John Horton}.} \bibinfo{year}{2013}\natexlab{}.
\newblock \showarticletitle{The Future of Crowd Work}. In \bibinfo{booktitle}{\emph{Proceedings of the 2013 {Conference} on {Computer} {Supported} {Cooperative} {Work}}} \emph{(\bibinfo{series}{{CSCW} '13})}. \bibinfo{publisher}{ACM}, \bibinfo{address}{New York, NY, USA}, \bibinfo{pages}{1301--1318}.
\newblock
\showISBNx{978-1-4503-1331-5}
\urldef\tempurl%
\url{https://doi.org/10.1145/2441776.2441923}
\showDOI{\tempurl}
\newblock
\shownote{00108}.


\bibitem[Knibbs(2024)]%
        {knibbs_congress_2024}
\bibfield{author}{\bibinfo{person}{Kate Knibbs}.} \bibinfo{year}{2024}\natexlab{}.
\newblock \bibinfo{title}{Congress {Wants} {Tech} {Companies} to {Pay} {Up} for {AI} {Training} {Data}}.
\newblock
\newblock
\showISSN{1059-1028}
\urldef\tempurl%
\url{https://www.wired.com/story/congress-senate-tech-companies-pay-ai-training-data/}
\showURL{%
\tempurl}


\bibitem[Kosmin and Waterman(1985)]%
        {kosminUseAndMisuseOfDJN1985}
\bibfield{author}{\bibinfo{person}{Barry~A. Kosmin} {and} \bibinfo{person}{Stanley Waterman}.} \bibinfo{year}{1985}\natexlab{}.
\newblock \showarticletitle{The Use and Misuse of Distinctive Jewish Names in Research on Jewish Populations}.
\newblock \bibinfo{journal}{\emph{Jewish Population Studies}}  \bibinfo{volume}{19} (\bibinfo{year}{1985}), \bibinfo{pages}{1--9}.
\newblock
\urldef\tempurl%
\url{https://api.semanticscholar.org/CorpusID:146336902}
\showURL{%
\tempurl}


\bibitem[Lampinen and Brown(2017)]%
        {lampinen2017market}
\bibfield{author}{\bibinfo{person}{Airi Lampinen} {and} \bibinfo{person}{Barry Brown}.} \bibinfo{year}{2017}\natexlab{}.
\newblock \showarticletitle{Market Design for HCI: Successes and Failures of Peer-to-Peer Exchange Platforms}. In \bibinfo{booktitle}{\emph{Proceedings of the 2017 CHI Conference on Human Factors in Computing Systems}} (Denver, Colorado, USA) \emph{(\bibinfo{series}{CHI '17})}. \bibinfo{publisher}{Association for Computing Machinery}, \bibinfo{address}{New York, NY, USA}, \bibinfo{pages}{4331–4343}.
\newblock
\showISBNx{9781450346559}
\urldef\tempurl%
\url{https://doi.org/10.1145/3025453.3025515}
\showDOI{\tempurl}


\bibitem[Lampinen et~al\mbox{.}(2018)]%
        {lampinen2018power}
\bibfield{author}{\bibinfo{person}{Airi Lampinen}, \bibinfo{person}{Christoph Lutz}, \bibinfo{person}{Gemma Newlands}, \bibinfo{person}{Ann Light}, {and} \bibinfo{person}{Nicole Immorlica}.} \bibinfo{year}{2018}\natexlab{}.
\newblock \showarticletitle{Power Struggles in the Digital Economy: Platforms, Workers, and Markets}. In \bibinfo{booktitle}{\emph{Companion of the 2018 ACM Conference on Computer Supported Cooperative Work and Social Computing}} \emph{(\bibinfo{series}{CSCW '18 Companion})}. \bibinfo{publisher}{Association for Computing Machinery}, \bibinfo{address}{New York, NY, USA}, \bibinfo{pages}{417–423}.
\newblock
\showISBNx{9781450360180}
\urldef\tempurl%
\url{https://doi.org/10.1145/3272973.3273004}
\showDOI{\tempurl}


\bibitem[Lazerwitz(1986)]%
        {lazerwitzSomeCommentsOnDJNs1986}
\bibfield{author}{\bibinfo{person}{Bernard Lazerwitz}.} \bibinfo{year}{1986}\natexlab{}.
\newblock \showarticletitle{Some comments on the use of distinctive Jewish names in surveys}.
\newblock \bibinfo{journal}{\emph{Contemporary Jewry}} \bibinfo{volume}{7}, \bibinfo{number}{1} (\bibinfo{date}{Jan} \bibinfo{year}{1986}), \bibinfo{pages}{83–91}.
\newblock
\showISSN{1876-5165}
\urldef\tempurl%
\url{https://doi.org/10.1007/BF02967944}
\showDOI{\tempurl}


\bibitem[Lee et~al\mbox{.}(2023a)]%
        {lee_ai_2023}
\bibfield{author}{\bibinfo{person}{Katherine Lee}, \bibinfo{person}{A.~Feder Cooper}, \bibinfo{person}{James Grimmelmann}, {and} \bibinfo{person}{Daphne Ippolito}.} \bibinfo{year}{2023}\natexlab{a}.
\newblock \showarticletitle{AI and Law: The Next Generation}. In \bibinfo{booktitle}{\emph{{GenLaw} '23 (at {ICML} '23)}}. \bibinfo{publisher}{The International Conference on Machine Learning (ICML)}, \bibinfo{address}{San Diego, CA, USA}, \bibinfo{pages}{1--21}.
\newblock
\urldef\tempurl%
\url{http://dx.doi.org/10.2139/ssrn.4580739}
\showURL{%
\tempurl}


\bibitem[Lee et~al\mbox{.}(2023b)]%
        {lee_generative_2023}
\bibfield{author}{\bibinfo{person}{Katherine Lee}, \bibinfo{person}{A.~Feder Cooper}, \bibinfo{person}{FatemehSadat Mireshghallah}, \bibinfo{person}{Madiha Zahrah}, \bibinfo{person}{James Grimmelmann}, \bibinfo{person}{David Mimno}, \bibinfo{person}{Deep Ganguli}, {and} \bibinfo{person}{Ludwig Schubert}.} \bibinfo{year}{2023}\natexlab{b}.
\newblock \bibinfo{booktitle}{\emph{Generative {AI} + {Law} ({GenLaw}) ’23}}.
\newblock The GenLaw Center.
\newblock
\urldef\tempurl%
\url{https://genlaw.github.io/}
\showURL{%
\tempurl}


\bibitem[Lee and Singh(2021)]%
        {lee2021landscape}
\bibfield{author}{\bibinfo{person}{Michelle Seng~Ah Lee} {and} \bibinfo{person}{Jat Singh}.} \bibinfo{year}{2021}\natexlab{}.
\newblock \showarticletitle{The Landscape and Gaps in Open Source Fairness Toolkits}. In \bibinfo{booktitle}{\emph{Proceedings of the 2021 CHI Conference on Human Factors in Computing Systems}} \emph{(\bibinfo{series}{CHI '21})}. \bibinfo{publisher}{Association for Computing Machinery}, \bibinfo{address}{New York, NY, USA}, Article \bibinfo{articleno}{699}, \bibinfo{numpages}{13}~pages.
\newblock
\showISBNx{9781450380966}
\urldef\tempurl%
\url{https://doi.org/10.1145/3411764.3445261}
\showDOI{\tempurl}


\bibitem[Lewis et~al\mbox{.}(2020)]%
        {lewisRetrievalAugmentedGenerationKnowledgeIntensive2020}
\bibfield{author}{\bibinfo{person}{Patrick Lewis}, \bibinfo{person}{Ethan Perez}, \bibinfo{person}{Aleksandra Piktus}, \bibinfo{person}{Fabio Petroni}, \bibinfo{person}{Vladimir Karpukhin}, \bibinfo{person}{Naman Goyal}, \bibinfo{person}{Heinrich Küttler}, \bibinfo{person}{Mike Lewis}, \bibinfo{person}{Wen-tau Yih}, \bibinfo{person}{Tim Rocktäschel}, \bibinfo{person}{Sebastian Riedel}, {and} \bibinfo{person}{Douwe Kiela}.} \bibinfo{year}{2020}\natexlab{}.
\newblock \showarticletitle{Retrieval-{Augmented} {Generation} for {Knowledge}-{Intensive} {NLP} {Tasks}}. In \bibinfo{booktitle}{\emph{Advances in {Neural} {Information} {Processing} {Systems}}}, Vol.~\bibinfo{volume}{33}. \bibinfo{publisher}{Curran Associates, Inc.}, \bibinfo{address}{San Dieco, CA, USA}, \bibinfo{pages}{9459--9474}.
\newblock
\urldef\tempurl%
\url{https://proceedings.neurips.cc/paper/2020/hash/6b493230205f780e1bc26945df7481e5-Abstract.html}
\showURL{%
\tempurl}


\bibitem[Li et~al\mbox{.}(2018)]%
        {li_out_2018}
\bibfield{author}{\bibinfo{person}{Hanlin Li}, \bibinfo{person}{Bodhisattva Alarcon}, \bibinfo{person}{Sara~Milkes Espinosa}, {and} \bibinfo{person}{Brent Hecht}.} \bibinfo{year}{2018}\natexlab{}.
\newblock \showarticletitle{Out of {Site}: {Empowering} a {New} {Approach} to {Online} {Boycotts}}. In \bibinfo{booktitle}{\emph{{CSCW} '18: 2018 {ACM} {Conference} on {Computer} {Supported} {Cooperative} {Work}}}. \bibinfo{publisher}{{ACM}}, \bibinfo{address}{{New York, NY, USA}}, \bibinfo{pages}{1--28}.
\newblock


\bibitem[Li et~al\mbox{.}(2023)]%
        {li2023dimensions}
\bibfield{author}{\bibinfo{person}{Hanlin Li}, \bibinfo{person}{Nicholas Vincent}, \bibinfo{person}{Stevie Chancellor}, {and} \bibinfo{person}{Brent Hecht}.} \bibinfo{year}{2023}\natexlab{}.
\newblock \showarticletitle{The Dimensions of Data Labor: A Road Map for Researchers, Activists, and Policymakers to Empower Data Producers}. In \bibinfo{booktitle}{\emph{Proceedings of the 2023 ACM Conference on Fairness, Accountability, and Transparency}}. \bibinfo{publisher}{{ACM}}, \bibinfo{address}{{New York, NY, USA}}, \bibinfo{pages}{1151--1161}.
\newblock


\bibitem[Liddy(2015)]%
        {liddy_urban_2015}
\bibfield{author}{\bibinfo{person}{Christian~D. Liddy}.} \bibinfo{year}{2015}\natexlab{}.
\newblock \showarticletitle{Urban {Enclosure} {Riots}: {Risings} of the {Commons} in {English} {Towns}, 1480–1525}.
\newblock \bibinfo{journal}{\emph{Past \& Present}} \bibinfo{volume}{226}, \bibinfo{number}{1} (\bibinfo{date}{Feb.} \bibinfo{year}{2015}), \bibinfo{pages}{41--77}.
\newblock
\showISSN{0031-2746}
\urldef\tempurl%
\url{https://doi.org/10.1093/pastj/gtu038}
\showDOI{\tempurl}


\bibitem[Longpre et~al\mbox{.}(2023)]%
        {longpredata}
\bibfield{author}{\bibinfo{person}{Shayne Longpre}, \bibinfo{person}{Robert Mahari}, \bibinfo{person}{Niklas Muennighoff}, \bibinfo{person}{Anthony Chen}, \bibinfo{person}{Kartik Perisetla}, \bibinfo{person}{William Brannon}, \bibinfo{person}{Jad Kabbara}, \bibinfo{person}{Luis Villa}, {and} \bibinfo{person}{Sara Hooker}.} \bibinfo{year}{2023}\natexlab{}.
\newblock \showarticletitle{The Data Provenance Project}. In \bibinfo{booktitle}{\emph{GenLaw Workshop at ICML}}. \bibinfo{publisher}{The International Conference on Machine Learning (ICML)}, \bibinfo{address}{San Diego, CA, USA}, \bibinfo{pages}{1--8}.
\newblock


\bibitem[Madaio et~al\mbox{.}(2020)]%
        {madaio2020codesign}
\bibfield{author}{\bibinfo{person}{Michael~A. Madaio}, \bibinfo{person}{Luke Stark}, \bibinfo{person}{Jennifer Wortman~Vaughan}, {and} \bibinfo{person}{Hanna Wallach}.} \bibinfo{year}{2020}\natexlab{}.
\newblock \showarticletitle{Co-Designing Checklists to Understand Organizational Challenges and Opportunities around Fairness in AI}. In \bibinfo{booktitle}{\emph{Proceedings of the 2020 CHI Conference on Human Factors in Computing Systems}} (Honolulu, HI, USA) \emph{(\bibinfo{series}{CHI '20})}. \bibinfo{publisher}{Association for Computing Machinery}, \bibinfo{address}{New York, NY, USA}, \bibinfo{pages}{1–14}.
\newblock
\showISBNx{9781450367080}
\urldef\tempurl%
\url{https://doi.org/10.1145/3313831.3376445}
\showDOI{\tempurl}


\bibitem[Marker et~al\mbox{.}(2021)]%
        {markerJewishCommunityStudies21stCentury}
\bibfield{author}{\bibinfo{person}{David~A. Marker}, \bibinfo{person}{Shelley Brock}, \bibinfo{person}{Darby Steiger}, \bibinfo{person}{Jill DeMatteis}, {and} \bibinfo{person}{Hanna Popick}.} \bibinfo{year}{2021}\natexlab{}.
\newblock \showarticletitle{Jewish Community Studies in the Twenty-First Century}.
\newblock \bibinfo{journal}{\emph{Contemporary Jewry}} \bibinfo{volume}{41}, \bibinfo{number}{2} (\bibinfo{date}{Jun} \bibinfo{year}{2021}), \bibinfo{pages}{349–368}.
\newblock
\showISSN{1876-5165}
\urldef\tempurl%
\url{https://doi.org/10.1007/s12397-021-09388-w}
\showDOI{\tempurl}


\bibitem[Mateos(2014)]%
        {mateosClassifyingEthnicityThroughNames2014}
\bibfield{author}{\bibinfo{person}{Pablo Mateos}.} \bibinfo{year}{2014}\natexlab{}.
\newblock \showarticletitle{Classifying Ethnicity Through People’s Names}. In \bibinfo{booktitle}{\emph{Names, Ethnicity and Populations}} \emph{(\bibinfo{series}{Advances in Spatial Science})}. \bibinfo{publisher}{Springer}, \bibinfo{address}{Berlin, Heidelberg}, \bibinfo{pages}{117–144}.
\newblock
\showISBNx{978-3-642-45412-7}
\urldef\tempurl%
\url{https://doi.org/10.1007/978-3-642-45413-4\_6}
\showDOI{\tempurl}


\bibitem[McCartan et~al\mbox{.}(2023)]%
        {mccartanEstimatingRacialDisparities2023}
\bibfield{author}{\bibinfo{person}{Cory McCartan}, \bibinfo{person}{Jacob Goldin}, \bibinfo{person}{Daniel~E. Ho}, {and} \bibinfo{person}{Kosuke Imai}.} \bibinfo{year}{2023}\natexlab{}.
\newblock \showarticletitle{Estimating Racial Disparities When Race is Not Observed}.
\newblock \bibinfo{journal}{\emph{arXiv preprint arXiv:2303.02580}}  \bibinfo{volume}{abs/2303.02580} (\bibinfo{year}{2023}), \bibinfo{pages}{1--29}.
\newblock
\showeprint{2303.02580}~[stat.AP]


\bibitem[Mehrabi et~al\mbox{.}(2021)]%
        {mehrabi2021survey}
\bibfield{author}{\bibinfo{person}{Ninareh Mehrabi}, \bibinfo{person}{Fred Morstatter}, \bibinfo{person}{Nripsuta Saxena}, \bibinfo{person}{Kristina Lerman}, {and} \bibinfo{person}{Aram Galstyan}.} \bibinfo{year}{2021}\natexlab{}.
\newblock \showarticletitle{A Survey on Bias and Fairness in Machine Learning}.
\newblock \bibinfo{journal}{\emph{ACM computing surveys (CSUR)}} \bibinfo{volume}{54}, \bibinfo{number}{6} (\bibinfo{year}{2021}), \bibinfo{pages}{1--35}.
\newblock
\urldef\tempurl%
\url{http://arxiv.org/abs/1908.09635}
\showURL{%
\tempurl}


\bibitem[Miller(2023)]%
        {miller_aitxt_2023}
\bibfield{author}{\bibinfo{person}{Cullen Miller}.} \bibinfo{year}{2023}\natexlab{}.
\newblock \bibinfo{title}{ai.txt: {A} new way for websites to set permissions for {AI}}.
\newblock
\newblock
\urldef\tempurl%
\url{https://spawning.substack.com/p/aitxt-a-new-way-for-websites-to-set}
\showURL{%
\tempurl}


\bibitem[Min et~al\mbox{.}(2023)]%
        {minSILOLanguageModels2023a}
\bibfield{author}{\bibinfo{person}{Sewon Min}, \bibinfo{person}{Suchin Gururangan}, \bibinfo{person}{Eric Wallace}, \bibinfo{person}{Hannaneh Hajishirzi}, \bibinfo{person}{Noah~A. Smith}, {and} \bibinfo{person}{Luke Zettlemoyer}.} \bibinfo{year}{2023}\natexlab{}.
\newblock \bibinfo{title}{{SILO} {Language} {Models}: {Isolating} {Legal} {Risk} {In} a {Nonparametric} {Datastore}}.
\newblock
\newblock
\urldef\tempurl%
\url{https://doi.org/10.48550/arXiv.2308.04430}
\showDOI{\tempurl}
\newblock
\shownote{arXiv:2308.04430 [cs]}.


\bibitem[Murphy(2016)]%
        {murphyMostLeastEducated}
\bibfield{author}{\bibinfo{person}{Caryle Murphy}.} \bibinfo{year}{2016}\natexlab{}.
\newblock \bibinfo{title}{The most and least educated {U}.{S}. religious groups}.
\newblock
\newblock
\urldef\tempurl%
\url{https://www.pewresearch.org/short-reads/2016/11/04/the-most-and-least-educated-u-s-religious-groups/}
\showURL{%
\tempurl}


\bibitem[Museum(2017)]%
        {holocaustMemorialMuseumAryanization}
\bibfield{author}{\bibinfo{person}{United States Holocaust~Memorial Museum}.} \bibinfo{year}{2017}\natexlab{}.
\newblock \bibinfo{title}{Aryanization}.
\newblock
\newblock
\urldef\tempurl%
\url{https://encyclopedia.ushmm.org/content/en/article/aryanization}
\showURL{%
\tempurl}


\bibitem[Museum(2023)]%
        {europeJewishPop}
\bibfield{author}{\bibinfo{person}{United States Holocaust~Memorial Museum}.} \bibinfo{year}{2023}\natexlab{}.
\newblock \bibinfo{title}{Jewish Population of Europe}.
\newblock
\newblock
\urldef\tempurl%
\url{https://encyclopedia.ushmm.org/content/en/gallery/jewish-population-of-europe}
\showURL{%
\tempurl}


\bibitem[Nations(2021)]%
        {unGlobalPopulation}
\bibfield{author}{\bibinfo{person}{United Nations}.} \bibinfo{year}{2021}\natexlab{}.
\newblock \bibinfo{title}{Global Population}.
\newblock
\newblock
\urldef\tempurl%
\url{https://www.un.org/en/global-issues/population}
\showURL{%
\tempurl}


\bibitem[NCBI(2023)]%
        {ncbi-website}
\bibfield{author}{\bibinfo{person}{NCBI}.} \bibinfo{year}{2023}\natexlab{}.
\newblock \bibinfo{title}{National Center for Biotechnology Information}.
\newblock
\newblock
\urldef\tempurl%
\url{https://www.ncbi.nlm.nih.gov/}
\showURL{%
\tempurl}


\bibitem[of~State(2020)]%
        {us_department_of_state_building_2020}
\bibfield{author}{\bibinfo{person}{U.S.~Department of State}.} \bibinfo{year}{2020}\natexlab{}.
\newblock \bibinfo{title}{Building Coalitions and Alliances - The Canary in the Coal Mine? The Need for Cooperation}.
\newblock
\newblock
\urldef\tempurl%
\url{https://www.youtube.com/watch?v=Ne3dGStTN8Q}
\showURL{%
\tempurl}


\bibitem[OpenAI(2023)]%
        {openaiGPT4TechnicalReport2023}
\bibfield{author}{\bibinfo{person}{OpenAI}.} \bibinfo{year}{2023}\natexlab{}.
\newblock \bibinfo{title}{{GPT}-4 {Technical} {Report}}.
\newblock
\newblock
\urldef\tempurl%
\url{https://doi.org/10.48550/arXiv.2303.08774}
\showDOI{\tempurl}
\newblock
\shownote{arXiv:2303.08774 [cs]}.


\bibitem[{OpenAI}(2023)]%
        {openai_openai_2018}
\bibfield{author}{\bibinfo{person}{{OpenAI}}.} \bibinfo{year}{2023}\natexlab{}.
\newblock \bibinfo{booktitle}{\emph{{OpenAI} Charter}}.
\newblock OpenAI.
\newblock
\urldef\tempurl%
\url{https://openai.com/charter}
\showURL{%
\tempurl}


\bibitem[Orlowski(2023)]%
        {orlowskiInternetOriginalSin2023}
\bibfield{author}{\bibinfo{person}{Andrew Orlowski}.} \bibinfo{year}{2023}\natexlab{}.
\newblock \bibinfo{title}{The internet’s ‘original sin’ means {AI} will be a nightmare}.
\newblock
\newblock
\showISSN{0307-1235}
\urldef\tempurl%
\url{https://www.telegraph.co.uk/business/2023/08/21/internets-original-sin-ai-nightmare/}
\showURL{%
\tempurl}


\bibitem[Oza and Lenharo(2023)]%
        {ozaHeLaSettlement}
\bibfield{author}{\bibinfo{person}{Anil Oza} {and} \bibinfo{person}{Mariana Lenharo}.} \bibinfo{year}{2023}\natexlab{}.
\newblock \bibinfo{title}{How the ‘groundbreaking’ Henrietta Lacks settlement could change research}.
\newblock
\newblock
\urldef\tempurl%
\url{https://doi.org/10.1038/d41586-023-02479-8}
\showDOI{\tempurl}


\bibitem[Parliament(2023)]%
        {ukParliamentEnclosure}
\bibfield{author}{\bibinfo{person}{UK Parliament}.} \bibinfo{year}{2023}\natexlab{}.
\newblock \bibinfo{title}{Enclosing the land}.
\newblock
\newblock
\urldef\tempurl%
\url{https://www.parliament.uk/about/living-heritage/transformingsociety/towncountry/landscape/overview/enclosingland/}
\showURL{%
\tempurl}


\bibitem[Paullada et~al\mbox{.}(2021)]%
        {paullada2021data}
\bibfield{author}{\bibinfo{person}{Amandalynne Paullada}, \bibinfo{person}{Inioluwa~Deborah Raji}, \bibinfo{person}{Emily~M Bender}, \bibinfo{person}{Emily Denton}, {and} \bibinfo{person}{Alex Hanna}.} \bibinfo{year}{2021}\natexlab{}.
\newblock \showarticletitle{Data and its (dis) contents: A survey of dataset development and use in machine learning research}.
\newblock \bibinfo{journal}{\emph{Patterns}} \bibinfo{volume}{2}, \bibinfo{number}{11} (\bibinfo{year}{2021}), \bibinfo{pages}{1--14}.
\newblock
\urldef\tempurl%
\url{https://doi.org/10.1016/j.patter.2021.100336}
\showURL{%
\tempurl}


\bibitem[Penedo et~al\mbox{.}(2023)]%
        {penedo2023refinedweb}
\bibfield{author}{\bibinfo{person}{Guilherme Penedo}, \bibinfo{person}{Quentin Malartic}, \bibinfo{person}{Daniel Hesslow}, \bibinfo{person}{Ruxandra Cojocaru}, \bibinfo{person}{Alessandro Cappelli}, \bibinfo{person}{Hamza Alobeidli}, \bibinfo{person}{Baptiste Pannier}, \bibinfo{person}{Ebtesam Almazrouei}, {and} \bibinfo{person}{Julien Launay}.} \bibinfo{year}{2023}\natexlab{}.
\newblock \showarticletitle{The RefinedWeb dataset for Falcon LLM: outperforming curated corpora with web data, and web data only}.
\newblock \bibinfo{journal}{\emph{arXiv preprint arXiv:2306.01116}}  \bibinfo{volume}{abs/2306.01116} (\bibinfo{year}{2023}), \bibinfo{pages}{1--32}.
\newblock
\showeprint[arxiv]{2306.01116}~[cs.CL]
\urldef\tempurl%
\url{https://arxiv.org/abs/2306.01116}
\showURL{%
\tempurl}


\bibitem[Peters and Davis(2023)]%
        {petersNewYorkTimes2023}
\bibfield{author}{\bibinfo{person}{Jay Peters} {and} \bibinfo{person}{Wes Davis}.} \bibinfo{year}{2023}\natexlab{}.
\newblock \bibinfo{title}{The {New} {York} {Times} blocks {OpenAI}’s web crawler}.
\newblock
\newblock
\urldef\tempurl%
\url{https://www.theverge.com/2023/8/21/23840705/new-york-times-openai-web-crawler-ai-gpt}
\showURL{%
\tempurl}


\bibitem[Phillips(2007)]%
        {phillipsNumberingTHeJews2007}
\bibfield{author}{\bibinfo{person}{Benjamin Phillips}.} \bibinfo{year}{2007}\natexlab{}.
\newblock \showarticletitle{Numbering the Jews: Evaluating and Improving Surveys of American Jews}.
\newblock \bibinfo{journal}{\emph{Brandeis University ProQuest University Publishing}}  \bibinfo{volume}{I} (\bibinfo{date}{Feb} \bibinfo{year}{2007}), \bibinfo{pages}{1--506}.
\newblock
\urldef\tempurl%
\url{https://www.semanticscholar.org/paper/Numbering-the-Jews%3A-Evaluating-and-Improving-of-Phillips/4fb989e9344020d0a3ca85caf27417e15417077a}
\showURL{%
\tempurl}


\bibitem[Press(2023)]%
        {associatedpressJamesPattersonMargaret2023}
\bibfield{author}{\bibinfo{person}{Associated Press}.} \bibinfo{year}{2023}\natexlab{}.
\newblock \bibinfo{title}{James {Patterson}, {Margaret} {Atwood} among thousands of writers urging {AI} companies to honor copyrights}.
\newblock
\newblock
\urldef\tempurl%
\url{https://apnews.com/article/patterson-atwood-ai-open-letter-f2c434694ed22a64bd09abbb5742c1e5}
\showURL{%
\tempurl}


\bibitem[Presser(2023)]%
        {archiveBooks3Metadata}
\bibfield{author}{\bibinfo{person}{Shawn Presser}.} \bibinfo{year}{2023}\natexlab{}.
\newblock \bibinfo{title}{Books3 Metadata}.
\newblock
\newblock
\newblock
\shownote{\url{https://web.archive.org/web/20230000000000*/https://battle.shawwn.com/books3-metadata.jsonl}, last accessed 11/30/2023}.


\bibitem[Reisner(2023)]%
        {atlanticReisner2023}
\bibfield{author}{\bibinfo{person}{Alex Reisner}.} \bibinfo{year}{2023}\natexlab{}.
\newblock \bibinfo{title}{Revealed: The Authors Whose Pirated Books Are Powering Generative AI}.
\newblock
\newblock
\urldef\tempurl%
\url{https://www.theatlantic.com/technology/archive/2023/08/books3-ai-meta-llama-pirated-books/675063/}
\showURL{%
\tempurl}


\bibitem[Roberts et~al\mbox{.}(2020)]%
        {raffel2020exploring}
\bibfield{author}{\bibinfo{person}{Adam Roberts}, \bibinfo{person}{Colin Raffel}, \bibinfo{person}{Katherine Lee}, \bibinfo{person}{Michael Matena}, \bibinfo{person}{Noam Shazeer}, \bibinfo{person}{Peter~J. Liu}, \bibinfo{person}{Sharan Narang}, \bibinfo{person}{Wei Li}, {and} \bibinfo{person}{Yanqi Zhou}.} \bibinfo{year}{2020}\natexlab{}.
\newblock \showarticletitle{Exploring the Limits of Transfer Learning with a Unified Text-to-Text Transformer}.
\newblock \bibinfo{journal}{\emph{The Journal of Machine Learning Research}} \bibinfo{volume}{21}, \bibinfo{number}{1} (\bibinfo{year}{2020}), \bibinfo{pages}{5485--5551}.
\newblock


\bibitem[Rosenwaike(1990)]%
        {rosenwaikeLeadingSurnames}
\bibfield{author}{\bibinfo{person}{Ira Rosenwaike}.} \bibinfo{year}{1990}\natexlab{}.
\newblock \showarticletitle{Leading Surnames Among American Jews}.
\newblock \bibinfo{journal}{\emph{Names}}  \bibinfo{volume}{38} (\bibinfo{date}{Jun} \bibinfo{year}{1990}), \bibinfo{pages}{31–38}.
\newblock
\showISSN{1756-2279}
\urldef\tempurl%
\url{https://doi.org/10.1179/nam.1990.38.1-2.31}
\showDOI{\tempurl}


\bibitem[Roth(2023)]%
        {rothAnotherGroupWriters2023}
\bibfield{author}{\bibinfo{person}{Emma Roth}.} \bibinfo{year}{2023}\natexlab{}.
\newblock \bibinfo{title}{Another group of writers is suing {OpenAI} over copyright claims}.
\newblock
\newblock
\urldef\tempurl%
\url{https://www.theverge.com/2023/9/11/23869145/writers-sue-openai-chatgpt-copyright-claims}
\showURL{%
\tempurl}


\bibitem[Samuelson(2023)]%
        {samuelsonGenerativeAIMeets2023}
\bibfield{author}{\bibinfo{person}{Pamela Samuelson}.} \bibinfo{year}{2023}\natexlab{}.
\newblock \showarticletitle{Generative {AI} meets copyright}.
\newblock \bibinfo{journal}{\emph{Science}} \bibinfo{volume}{381}, \bibinfo{number}{6654} (\bibinfo{date}{July} \bibinfo{year}{2023}), \bibinfo{pages}{158--161}.
\newblock
\urldef\tempurl%
\url{https://doi.org/10.1126/science.adi0656}
\showDOI{\tempurl}
\newblock
\shownote{Publisher: American Association for the Advancement of Science}.


\bibitem[Santos(2024)]%
        {santosChatGPTAIReplacing2024}
\bibfield{author}{\bibinfo{person}{Melissa Santos}.} \bibinfo{year}{2024}\natexlab{}.
\newblock \bibinfo{title}{{ChatGPT} and {AI} replacing jobs is a worry for workers, per {WSU} survey}.
\newblock
\newblock
\urldef\tempurl%
\url{https://www.axios.com/local/seattle/2024/02/09/chat-gpt-ai-workers-replace-employees}
\showURL{%
\tempurl}


\bibitem[Sarna(2019)]%
        {sarnaAmericanJewishPopulationEstimates}
\bibfield{author}{\bibinfo{person}{Jonathan~D. Sarna}.} \bibinfo{year}{2019}\natexlab{}.
\newblock \bibinfo{booktitle}{\emph{Appendix: American Jewish Population Estimates, 1660–2015}}.
\newblock \bibinfo{publisher}{Yale University Press}, \bibinfo{address}{New Haven, CT, USA}, \bibinfo{pages}{391–392}.
\newblock
\showISBNx{978-0-300-19039-7}
\urldef\tempurl%
\url{https://doi.org/10.2307/j.ctvhrczf4.13}
\showDOI{\tempurl}


\bibitem[Saxe et~al\mbox{.}(2021)]%
        {ajppBrandeisReport}
\bibfield{author}{\bibinfo{person}{Leonard Saxe}, \bibinfo{person}{Daniel Parmer}, \bibinfo{person}{Elizabeth Tighe}, \bibinfo{person}{Raquel~Magidin de Kramer}, \bibinfo{person}{Daniel Kallista}, \bibinfo{person}{Daniel Nussbaum}, \bibinfo{person}{Xajavion Seabrum}, {and} \bibinfo{person}{Joshua Mandell}.} \bibinfo{year}{2021}\natexlab{}.
\newblock \bibinfo{booktitle}{\emph{American Jewish Population Estimates 2020: Summary \& Highlights}}.
\newblock Brandeis University.
\newblock


\bibitem[Schaul et~al\mbox{.}(2023)]%
        {schaulSecretListWebsites}
\bibfield{author}{\bibinfo{person}{Kevin Schaul}, \bibinfo{person}{Szu~Yu Chen}, {and} \bibinfo{person}{Nitasha Tiku}.} \bibinfo{year}{2023}\natexlab{}.
\newblock \bibinfo{title}{Inside the secret list of websites that make {AI} like {ChatGPT} sound smart}.
\newblock
\newblock
\urldef\tempurl%
\url{https://www.washingtonpost.com/technology/interactive/2023/ai-chatbot-learning/}
\showURL{%
\tempurl}


\bibitem[Senftleben(2023)]%
        {senftlebenGenerativeAIAuthor2023}
\bibfield{author}{\bibinfo{person}{Martin Senftleben}.} \bibinfo{year}{2023}\natexlab{}.
\newblock \showarticletitle{Generative {AI} and {Author} {Remuneration}}.
\newblock \bibinfo{journal}{\emph{IIC - International Review of Intellectual Property and Competition Law}} \bibinfo{volume}{54}, \bibinfo{number}{10} (\bibinfo{date}{Nov.} \bibinfo{year}{2023}), \bibinfo{pages}{1535--1560}.
\newblock
\showISSN{2195-0237}
\urldef\tempurl%
\url{https://doi.org/10.1007/s40319-023-01399-4}
\showDOI{\tempurl}


\bibitem[Service(2023)]%
        {congressionallegalserviceGenerativeArtificialIntelligence2023}
\bibfield{author}{\bibinfo{person}{Congressional~Legal Service}.} \bibinfo{year}{2023}\natexlab{}.
\newblock \bibinfo{booktitle}{\emph{Generative {Artificial} {Intelligence} and {Copyright} {Law}}}.
\newblock \bibinfo{type}{{T}echnical {R}eport}. \bibinfo{institution}{Congressional Legal Service}.
\newblock
\urldef\tempurl%
\url{https://crsreports.congress.gov/product/pdf/LSB/LSB10922}
\showURL{%
\tempurl}


\bibitem[Sheskin(1998)]%
        {sheskinMiami1998}
\bibfield{author}{\bibinfo{person}{Ira~M. Sheskin}.} \bibinfo{year}{1998}\natexlab{}.
\newblock \showarticletitle{A Methodology for Examining the Changing Size and Spatial Distribution of a Jewish Population: A Miami Case Study}.
\newblock \bibinfo{journal}{\emph{Shofar: An Interdisciplinary Journal of Jewish Studies}} \bibinfo{volume}{17}, \bibinfo{number}{1} (\bibinfo{year}{1998}), \bibinfo{pages}{97–116}.
\newblock
\showISSN{0882-8539}
\urldef\tempurl%
\url{https://doi.org/10.1353/sho.1998.0041}
\showURL{%
\tempurl}


\bibitem[Sheskin(2016)]%
        {sheskinGoodPractices2016}
\bibfield{author}{\bibinfo{person}{Ira~M. Sheskin}.} \bibinfo{year}{2016}\natexlab{}.
\newblock \showarticletitle{Good Practices in Local Jewish Community Studies}.
\newblock \bibinfo{journal}{\emph{Contemporary Jewry}} \bibinfo{volume}{36}, \bibinfo{number}{3} (\bibinfo{date}{Oct} \bibinfo{year}{2016}), \bibinfo{pages}{319–341}.
\newblock
\showISSN{1876-5165}
\urldef\tempurl%
\url{https://doi.org/10.1007/s12397-016-9184-3}
\showDOI{\tempurl}


\bibitem[Sheskin and Dashefsky(2012)]%
        {sheskinJewishPopulation2012}
\bibfield{author}{\bibinfo{person}{Ira~M. Sheskin} {and} \bibinfo{person}{Arnold Dashefsky}.} \bibinfo{year}{2012}\natexlab{}.
\newblock \showarticletitle{Jewish Population in the United States, 2012}.
\newblock \bibinfo{journal}{\emph{The American Jewish Year Book}}  \bibinfo{volume}{109/112} (\bibinfo{year}{2012}), \bibinfo{pages}{143–211}.
\newblock
\showISSN{0065-8987}


\bibitem[Sheskin and Dashefsky(2023)]%
        {allAJYB}
\bibfield{editor}{\bibinfo{person}{Ira~M. Sheskin} {and} \bibinfo{person}{Arnold Dashefsky}} (Eds.). \bibinfo{year}{2012-2023}\natexlab{}.
\newblock \bibinfo{booktitle}{\emph{American Jewish Year Book (series)}}.
\newblock \bibinfo{publisher}{Jewish Publication Society; American Jewish Committee}, \bibinfo{address}{USA}.
\newblock
\urldef\tempurl%
\url{https://www.springer.com/series/11193}
\showURL{%
\tempurl}


\bibitem[Shi et~al\mbox{.}(2023)]%
        {shiDetectingPretrainingData2023}
\bibfield{author}{\bibinfo{person}{Weijia Shi}, \bibinfo{person}{Anirudh Ajith}, \bibinfo{person}{Mengzhou Xia}, \bibinfo{person}{Yangsibo Huang}, \bibinfo{person}{Daogao Liu}, \bibinfo{person}{Terra Blevins}, \bibinfo{person}{Danqi Chen}, {and} \bibinfo{person}{Luke Zettlemoyer}.} \bibinfo{year}{2023}\natexlab{}.
\newblock \bibinfo{title}{Detecting {Pretraining} {Data} from {Large} {Language} {Models}}.
\newblock
\newblock
\urldef\tempurl%
\url{https://doi.org/10.48550/arXiv.2310.16789}
\showDOI{\tempurl}
\newblock
\shownote{arXiv:2310.16789 [cs]}.


\bibitem[Smith(2024)]%
        {smithAIStartingThreaten2024}
\bibfield{author}{\bibinfo{person}{Ray~A. Smith}.} \bibinfo{year}{2024}\natexlab{}.
\newblock \bibinfo{title}{{AI} {Is} {Starting} to {Threaten} {White}-{Collar} {Jobs}. {Few} {Industries} {Are} {Immune}.}
\newblock
\newblock
\urldef\tempurl%
\url{https://www.wsj.com/lifestyle/careers/ai-is-starting-to-threaten-white-collar-jobs-few-industries-are-immune-9cdbcb90}
\showURL{%
\tempurl}
\newblock
\shownote{Section: Management}.


\bibitem[Tao et~al\mbox{.}(2021)]%
        {taoTrendVirtualHybrid2021}
\bibfield{author}{\bibinfo{person}{Yanqiu Tao}, \bibinfo{person}{Debbie Steckel}, \bibinfo{person}{Jiří~Jaromír Klemeš}, {and} \bibinfo{person}{Fengqi You}.} \bibinfo{year}{2021}\natexlab{}.
\newblock \showarticletitle{Trend towards virtual and hybrid conferences may be an effective climate change mitigation strategy}.
\newblock \bibinfo{journal}{\emph{Nature Communications}} \bibinfo{volume}{12}, \bibinfo{number}{1} (\bibinfo{date}{Dec.} \bibinfo{year}{2021}), \bibinfo{pages}{7324}.
\newblock
\showISSN{2041-1723}
\urldef\tempurl%
\url{https://doi.org/10.1038/s41467-021-27251-2}
\showDOI{\tempurl}
\newblock
\shownote{Number: 1 Publisher: Nature Publishing Group}.


\bibitem[Teller(2010)]%
        {tellerYivoEconomicLife}
\bibfield{author}{\bibinfo{person}{Adam Teller}.} \bibinfo{year}{2010}\natexlab{}.
\newblock \bibinfo{title}{Economic Life}.
\newblock
\newblock
\urldef\tempurl%
\url{https://yivoencyclopedia.org/article.aspx/economic_life}
\showURL{%
\tempurl}


\bibitem[Touvron et~al\mbox{.}(2023a)]%
        {touvronLLaMAOpenEfficient2023}
\bibfield{author}{\bibinfo{person}{Hugo Touvron}, \bibinfo{person}{Thibaut Lavril}, \bibinfo{person}{Gautier Izacard}, \bibinfo{person}{Xavier Martinet}, \bibinfo{person}{Marie-Anne Lachaux}, \bibinfo{person}{Timothée Lacroix}, \bibinfo{person}{Baptiste Rozière}, \bibinfo{person}{Naman Goyal}, \bibinfo{person}{Eric Hambro}, \bibinfo{person}{Faisal Azhar}, \bibinfo{person}{Aurelien Rodriguez}, \bibinfo{person}{Armand Joulin}, \bibinfo{person}{Edouard Grave}, {and} \bibinfo{person}{Guillaume Lample}.} \bibinfo{year}{2023}\natexlab{a}.
\newblock \bibinfo{title}{{LLaMA}: {Open} and {Efficient} {Foundation} {Language} {Models}}.
\newblock
\newblock
\urldef\tempurl%
\url{https://doi.org/10.48550/arXiv.2302.13971}
\showDOI{\tempurl}


\bibitem[Touvron et~al\mbox{.}(2023b)]%
        {touvron2023llama}
\bibfield{author}{\bibinfo{person}{Hugo Touvron}, \bibinfo{person}{Louis Martin}, \bibinfo{person}{Kevin Stone}, \bibinfo{person}{Peter Albert}, \bibinfo{person}{Amjad Almahairi}, \bibinfo{person}{Yasmine Babaei}, \bibinfo{person}{Nikolay Bashlykov}, \bibinfo{person}{Soumya Batra}, \bibinfo{person}{Prajjwal Bhargava}, \bibinfo{person}{Shruti Bhosale}, {et~al\mbox{.}}} \bibinfo{year}{2023}\natexlab{b}.
\newblock \showarticletitle{Llama 2: Open foundation and fine-tuned chat models}.
\newblock \bibinfo{journal}{\emph{arXiv preprint arXiv:2307.09288}}  \bibinfo{volume}{abs/2307.09288} (\bibinfo{year}{2023}), \bibinfo{pages}{1--77}.
\newblock
\urldef\tempurl%
\url{https://arxiv.org/abs/2307.09288}
\showURL{%
\tempurl}


\bibitem[van Berkel et~al\mbox{.}(2021)]%
        {vanberkel2021presentation}
\bibfield{author}{\bibinfo{person}{Niels van Berkel}, \bibinfo{person}{Jorge Goncalves}, \bibinfo{person}{Daniel Russo}, \bibinfo{person}{Simo Hosio}, {and} \bibinfo{person}{Mikael~B. Skov}.} \bibinfo{year}{2021}\natexlab{}.
\newblock \showarticletitle{Effect of Information Presentation on Fairness Perceptions of Machine Learning Predictors}. In \bibinfo{booktitle}{\emph{Proceedings of the 2021 CHI Conference on Human Factors in Computing Systems}} \emph{(\bibinfo{series}{CHI '21})}. \bibinfo{publisher}{Association for Computing Machinery}, \bibinfo{address}{New York, NY, USA}, Article \bibinfo{articleno}{245}, \bibinfo{numpages}{13}~pages.
\newblock
\showISBNx{9781450380966}
\urldef\tempurl%
\url{https://doi.org/10.1145/3411764.3445365}
\showDOI{\tempurl}


\bibitem[Vincent(2023)]%
        {vincentGettyImagesSues2023}
\bibfield{author}{\bibinfo{person}{James Vincent}.} \bibinfo{year}{2023}\natexlab{}.
\newblock \bibinfo{title}{Getty {Images} sues {AI} art generator {Stable} {Diffusion} in the {US} for copyright infringement}.
\newblock
\newblock
\urldef\tempurl%
\url{https://www.theverge.com/2023/2/6/23587393/ai-art-copyright-lawsuit-getty-images-stable-diffusion}
\showURL{%
\tempurl}


\bibitem[Vincent and Hecht(2023)]%
        {vincent_sharing_2023}
\bibfield{author}{\bibinfo{person}{Nicholas Vincent} {and} \bibinfo{person}{Brent Hecht}.} \bibinfo{year}{2023}\natexlab{}.
\newblock \bibinfo{title}{Sharing the {Winnings} of {AI} with {Data} {Dividends}: {Challenges} with ``{Meritocratic}'' {Data} {Valuation}}.
\newblock
\newblock


\bibitem[Vincent et~al\mbox{.}(2019)]%
        {vincent_data_2019}
\bibfield{author}{\bibinfo{person}{Nicholas Vincent}, \bibinfo{person}{Brent Hecht}, {and} \bibinfo{person}{Shilad Sen}.} \bibinfo{year}{2019}\natexlab{}.
\newblock \showarticletitle{“Data Strikes”: Evaluating the Effectiveness of a New Form of Collective Action Against Technology Companies}. In \bibinfo{booktitle}{\emph{The World Wide Web Conference}} (San Francisco, CA, USA) \emph{(\bibinfo{series}{WWW '19})}. \bibinfo{publisher}{Association for Computing Machinery}, \bibinfo{address}{New York, NY, USA}, \bibinfo{pages}{1931–1943}.
\newblock
\showISBNx{9781450366748}
\urldef\tempurl%
\url{https://doi.org/10.1145/3308558.3313742}
\showDOI{\tempurl}


\bibitem[Vincent and Li(2023)]%
        {vincent_chatgpt_2023}
\bibfield{author}{\bibinfo{person}{Nick Vincent} {and} \bibinfo{person}{Hanlin Li}.} \bibinfo{year}{2023}\natexlab{}.
\newblock \bibinfo{title}{{ChatGPT} Stole Your Work. So What Are You Going to Do?}
\newblock
\newblock
\showISSN{1059-1028}
\urldef\tempurl%
\url{https://www.wired.com/story/chatgpt-generative-artificial-intelligence-regulation/}
\showURL{%
\tempurl}


\bibitem[Vincent(2020)]%
        {vincent_dont_2020}
\bibfield{author}{\bibinfo{person}{Nicholas~M. Vincent}.} \bibinfo{year}{2020}\natexlab{}.
\newblock \bibinfo{booktitle}{\emph{Don’t give {OpenAI} all the credit for {GPT}-3: You might have helped create the latest “astonishing” advance in {AI} too}}.
\newblock People, Space, and Algorithms Research Group.
\newblock
\urldef\tempurl%
\url{https://www.psagroup.org/blogposts/62}
\showURL{%
\tempurl}


\bibitem[Wachs et~al\mbox{.}(2022)]%
        {wachsGeographyOfGitHub}
\bibfield{author}{\bibinfo{person}{Johannes Wachs}, \bibinfo{person}{Mariusz Nitecki}, \bibinfo{person}{William Schueller}, {and} \bibinfo{person}{Axel Polleres}.} \bibinfo{year}{2022}\natexlab{}.
\newblock \showarticletitle{The Geography of Open Source Software: Evidence from GitHub}.
\newblock \bibinfo{journal}{\emph{Technological Forecasting and Social Change}}  \bibinfo{volume}{176} (\bibinfo{date}{Mar} \bibinfo{year}{2022}), \bibinfo{pages}{121478}.
\newblock
\showISSN{0040-1625}
\urldef\tempurl%
\url{https://doi.org/10.1016/j.techfore.2022.121478}
\showDOI{\tempurl}


\bibitem[Wikipedia(2023)]%
        {noauthor_rene_2023}
\bibfield{author}{\bibinfo{person}{Wikipedia}.} \bibinfo{year}{2023}\natexlab{}.
\newblock \bibinfo{booktitle}{\emph{René Carmille}}.
\newblock Wikipedia.
\newblock
\urldef\tempurl%
\url{https://fr.wikipedia.org/w/index.php?title=Ren%C3%A9_Carmille&oldid=210288825}
\showURL{%
\tempurl}
\newblock
\shownote{Page Version {ID}: 210288825}.


\bibitem[Xia et~al\mbox{.}(2022)]%
        {xiaDataStationDelegated2022}
\bibfield{author}{\bibinfo{person}{Siyuan Xia}, \bibinfo{person}{Zhiru Zhu}, \bibinfo{person}{Chris Zhu}, \bibinfo{person}{Jinjin Zhao}, \bibinfo{person}{Kyle Chard}, \bibinfo{person}{Aaron~J. Elmore}, \bibinfo{person}{Ian Foster}, \bibinfo{person}{Michael Franklin}, \bibinfo{person}{Sanjay Krishnan}, {and} \bibinfo{person}{Raul~Castro Fernandez}.} \bibinfo{year}{2022}\natexlab{}.
\newblock \showarticletitle{Data station: delegated, trustworthy, and auditable computation to enable data-sharing consortia with a data escrow}.
\newblock \bibinfo{journal}{\emph{Proceedings of the VLDB Endowment}} \bibinfo{volume}{15}, \bibinfo{number}{11} (\bibinfo{date}{July} \bibinfo{year}{2022}), \bibinfo{pages}{3172--3185}.
\newblock
\showISSN{2150-8097}
\urldef\tempurl%
\url{https://doi.org/10.14778/3551793.3551861}
\showDOI{\tempurl}


\bibitem[Yilmaz et~al\mbox{.}(2023)]%
        {yilmazAIDrivenLaborSubstitution2023}
\bibfield{author}{\bibinfo{person}{Erdem~Dogukan Yilmaz}, \bibinfo{person}{Ivana Naumovska}, {and} \bibinfo{person}{Vikas~A. Aggarwal}.} \bibinfo{year}{2023}\natexlab{}.
\newblock \bibinfo{title}{{AI}-{Driven} {Labor} {Substitution}: {Evidence} from {Google} {Translate} and {ChatGPT}}.
\newblock
\newblock
\urldef\tempurl%
\url{https://doi.org/10.2139/ssrn.4400516}
\showDOI{\tempurl}


\bibitem[Zamani et~al\mbox{.}(2022)]%
        {zamaniRetrievalEnhancedMachineLearning2022}
\bibfield{author}{\bibinfo{person}{Hamed Zamani}, \bibinfo{person}{Fernando Diaz}, \bibinfo{person}{Mostafa Dehghani}, \bibinfo{person}{Donald Metzler}, {and} \bibinfo{person}{Michael Bendersky}.} \bibinfo{year}{2022}\natexlab{}.
\newblock \showarticletitle{Retrieval-{Enhanced} {Machine} {Learning}}. In \bibinfo{booktitle}{\emph{Proceedings of the 45th {International} {ACM} {SIGIR} {Conference} on {Research} and {Development} in {Information} {Retrieval}}}. \bibinfo{publisher}{Special Interest Group on Information Retrieval (SIGIR)}, \bibinfo{address}{NYC, NY, USA}, \bibinfo{pages}{2875--2886}.
\newblock
\urldef\tempurl%
\url{https://doi.org/10.1145/3477495.3531722}
\showDOI{\tempurl}
\newblock
\shownote{arXiv:2205.01230 [cs]}.


\bibitem[Zarifhonarvar(2023)]%
        {zarifhonarvarEconomicsChatGPTLabor2023}
\bibfield{author}{\bibinfo{person}{Ali Zarifhonarvar}.} \bibinfo{year}{2023}\natexlab{}.
\newblock \bibinfo{title}{Economics of {ChatGPT}: {A} {Labor} {Market} {View} on the {Occupational} {Impact} of {Artificial} {Intelligence}}.
\newblock
\newblock
\urldef\tempurl%
\url{https://doi.org/10.2139/ssrn.4350925}
\showDOI{\tempurl}


\bibitem[Zeng et~al\mbox{.}(2023)]%
        {Zeng2023LargeLM}
\bibfield{author}{\bibinfo{person}{Fanlong Zeng}, \bibinfo{person}{Wensheng Gan}, \bibinfo{person}{Yongheng Wang}, \bibinfo{person}{Ning Liu}, {and} \bibinfo{person}{Philip~S. Yu}.} \bibinfo{year}{2023}\natexlab{}.
\newblock \showarticletitle{Large Language Models for Robotics: A Survey}.
\newblock \bibinfo{journal}{\emph{ArXiv}}  \bibinfo{volume}{abs/2311.07226} (\bibinfo{year}{2023}), \bibinfo{pages}{1--19}.
\newblock
\urldef\tempurl%
\url{https://api.semanticscholar.org/CorpusID:265149884}
\showURL{%
\tempurl}


\end{thebibliography}

\appendix
\section{Appendix}
This appendix covers three topics across five sections:
\begin{itemize}
    \item Additional robustness checks and analyses we performed to increase confidence in our methods (\ref{app:how-wrong}, \ref{app:us-estimation}, \ref{app:dataset-age-of-freelaw}).
    \item More information on the history and use of DJNs (\ref{app:djns}).
    \item A supplementary analysis that measures IP dispossession by industry. We include it here as both a robustness check (the results seem to support our conclusions) and as a topic for future research (\ref{app:ip-disp-by-industry}).
\end{itemize}

\subsection{How Wrong Would Our Estimates Have To Be To Achieve No Impact}
\label{app:how-wrong}
In the main body of this study, we identify a reasonable range for each parameter to determine lower and upper bounds for each estimate. Here, we check how robust the results are to even more conservative parameter values. All of these values are massive departures from the best available data discussed in the main body of our paper; we include this section to validate the robustness of our claims. To do this, we calculated what value each parameter would have to have in order to indicate no over-representation. We held all parameters other than the one under investigation constant at the original lower bound estimates. We found that in order for the weighted relative dispossession magnitude to be $\sim$1 X (ie: no overrepresentation), the following would have to be true:
\begin{itemize}
    \item 32.6\% or fewer of people with DJNs are Jewish (the remaining 67.4\% are non-Jews).
    \item 27.4\% or more of the U.S. Jewish population have DJNs.
    \item 5.9\% of the U.S. population is Jewish.
\end{itemize}

These precise values for each parameter are in Tables \ref{fig:PIPDE-how-wrong-precision}-\ref{fig:PIPDE-how-wrong-us-pop}.

\begin{table}[ht]
    \centering
    \begin{tabular}{ll}
    \hline
    \textbf{Parameter} & 
    \textbf{Estimates} \\
    \hline
    
    \% precision of DJNs                 & \textbf{32.6\%} \\
    \% coverage of DJNs                  & 11.18\%  \\
    \% of US population that is Jewish   & 2.4\%  \\
    \\
    \textbf{Relative Dispossession Magnitude}  \\
    Total                                & 1.17 X  \\
    Weighted Total                       & 1.00 X\\ \\

    \end{tabular}
    \caption{All parameters other than precision are held constant at lower bound estimates. With precision at 32.6\%, the weighted total relative dispossession magnitude would be 1.00 X.}
    \label{fig:PIPDE-how-wrong-precision}
\end{table}

\begin{table}[ht]
    \centering
    \begin{tabular}{ll}
    \hline
    \textbf{Parameter} & 
    \textbf{Estimates} \\
    \hline
    
    \% precision of DJNs                 & 80\% \\
    \% coverage of DJNs                  & \textbf{27.4\%}  \\
    \% of US population that is Jewish   & 2.4\%  \\
    \\
    \textbf{Relative Dispossession Magnitude}  \\
    Total                                & 1.17 X  \\
    Weighted Total                       & 1.00 X\\ \\

    \end{tabular}
    \caption{All parameters other than coverage are held constant at lower bound estimates. With coverage at 27.4\%, the weighted total relative dispossession magnitude would be 1.00 X.}
    \label{fig:PIPDE-how-wrong-coverage}
\end{table}

\begin{table}[ht]
    \centering
    \begin{tabular}{ll}
    \hline
    \textbf{Parameter} & 
    \textbf{Estimates} \\
    \hline
    
    \% precision of DJNs                 & 80\% \\
    \% coverage of DJNs                  & 11.18\%  \\
    \% of US population that is Jewish   & \textbf{5.9\%}  \\
    \\
    \textbf{Relative Dispossession Magnitude}  \\
    Total                                & 1.16 X  \\
    Weighted Total                       & 1.00 X\\ \\

    \end{tabular}
    \caption{All parameters other than the \% of the U.S. population that is Jewish are held constant. With U.S. Jews at 5.9\% of the population, the weighted total relative dispossession magnitude would be 1.00 X.}
    \label{fig:PIPDE-how-wrong-us-pop}
\end{table}

\subsection{Estimation of U.S. Representation In Dataset}
\label{app:us-estimation}
Here, we discuss how we investigated the extent to which each dataset included content from people based in the U.S. This helped us increase our confidence in our estimate of the relative representation of Jewish Americans relative other Americans. An alternative approach in future work might be to consider representative purely at the global level.
\subsubsection{Pubmed Central}
\label{app:pubmed-central}
We used journal countries of publication~\cite{ncbi-website} as a proxy for author countries to estimate an expected 0.6\% of Jewish authorship as compared to the 1.95-2.21\% we see (Table \ref{fig:pmc-countries}).

\begin{table}[H]
    \centering
    \begin{tabular}{|l|ll|}
    \hline
    \multicolumn{1}{|l|}{\multirow{2}{*}{\textbf{\begin{tabular}[c]{@{}l@{}}Journals\\ in PMC\end{tabular}}}} & \multicolumn{2}{|l|}{\textbf{\begin{tabular}[c]{@{}l@{}}Journals by Country \\ of Publication\end{tabular}}} \\ 
    \cline{2-3} 
    \multicolumn{1}{|l|}{}     & \textbf{U.S.}   & \multicolumn{1}{l|}{\textbf{Other}}       \\ \hline
    3,499       & 1,000        & 2,499 \\ \hline 
    \end{tabular}
    \caption{Publication countries of PMC journals, intended as a rough estimate of percentage of the dataset that is U.S. based or affiliated.}
    \label{fig:pmc-countries}
\end{table}

\subsubsection{ArXiv}
\label{app:arxiv}

We estimated the number of U.S.\ authors who had submitted work to ArXiv. We scoped U.S.\ authorship to those who work at U.S.\ institutions, as they are most likely U.S.\ residents. Using data from the U.S.\ Department of Education's National Center for Education Statistics (NCES),\footnote{\url{https://nces.ed.gov/}} which tracks educational institutions that accept federal funding, we identified U.S.\ higher education institutions. We then manually inspected all ArXiv institutions containing the word ``hospital'' or ``medical'' to find research originating in U.S.\ hospital systems. We then sorted the remaining institutions by highest number of submissions and added U.S.\ research institutes (e.g., NIST, Los Alamos National Laboratory) and companies (e.g., Intel, IBM) to our list, until no U.S.-based institutions remained in the top 500 institutions. From this list, we calculated the number of U.S.\ submissions. This is likely to be an undercount, as there are over 6200 educational institutions and there are likely formatting inconsistencies between ArXiv and NCES that will have unintentionally excluded U.S.\ institutions.

\subsection{Robustness Check for Document Age in FreeLaw}
\label{app:dataset-age-of-freelaw}
We considered constructing a weighted average of Jewish population over time based on the publication dates of documents in the dataset. When we used FreeLaw as a case study (because it is the dataset with the oldest records, covers the greatest time span, and is entirely U.S.-based), we found that it barely affected our expected Jewish population percentage. Because documents in the rest of the sources are more recent, and the Jewish population has stayed somewhat constant over the past 10-15 years (when many of the documents across datasets were produced), we decided \textit{not} to stray from the naive approach to account for document age. We note here that if a significant number of documents in a source were published between 2000-2010 we'd expect to see a slightly lower percentage of Jews, and if a significant number of documents were published between 1920-1970 we'd expect to see a slightly higher percentage.

\subsection{Additional Detail On the Distinctive Jewish Names-Based Methodology}
\label{app:djns}
As summarized briefly in Section \ref{related:djn}, DJNs have a long history in the world of Jewish demography. The original concept is attributed to Kohs, who found that the most common surnames on Jewish Federation membership lists represented a significant proportion of overall membership ~\cite{himmelfarbSamplingByEthnicSurnames}. Later surveys in the 1960s-80s confirmed that, depending on the list used, the national proportion of Jews with DJNs remained roughly constant at $\sim$11-12\%. Since then, DJNs have been used in numerous studies, including (a) as the basis for the sampling frame (or more likely, as the basis for one of several sampling frames) for local Jewish community studies; (b) to measure the change in size of a Jewish community over time; and (c) to estimate the overall size of a Jewish community within a larger local population (e.g., the number of Jews on a college campus ~\cite{sheskinCollegeCampus2013}).\footnote{See the ``United States Jewish Population'' chapters of the American Jewish Yearbook series for a detailed account of such studies ~\cite{allAJYB, dellaPergolaAJYB2021}.} Any potential use of DJNs should be measured on at least two axes---coverage and representativeness---which we elaborate on below.

\subsubsection{Coverage}
Research that uses DJN-based lists as a sole sampling frame by definition can only cover the percentage of the population with a Jewish surname. In local community studies, this puts a strong cap on possible participants ~\cite{dutwinEverythingToConsider2016}. As the current study does not directly survey a sub-population, coverage in that sense is not an issue so long as we can calculate the proportion of the U.S.\ Jewish community that our sample captures. In the earliest studies of DJNs, researchers estimated that between 11--12\% of the national Jewish population had one of $\sim$35 surnames (~\cite{himmelfarbSamplingByEthnicSurnames, kosminUseAndMisuseOfDJN1985, lazerwitzSomeCommentsOnDJNs1986}). However, they caution that these numbers vary significantly across local subsets of the Jewish community, so smaller studies should be wary of applying the 11--12\% figure ``unless they have prior knowledge about the actual size of the proportion of the population that DJN persons constitute and the stability of that proportion over time.'' ~\cite{himmelfarbFurtherComments1876}.
    
These numbers have remained largely consistent over time for large U.S.\ Jewish communities. In 2007, ~\citeauthor{phillipsNumberingTHeJews2007} found that 12.4\% of Greater Boston Area Jews had a DJN, and 91.8\% of people with a DJN were Jewish. In a 2012 review of local area studies, ~\citeauthor{sheskinJewishPopulation2012} noted, ``the fact that about 8--12\% of American Jews, despite rising intermarriage, continue to have one of 36 Distinctive Jewish Names... facilitates making reasonable estimates of the Jewish population.'' ~\cite{sheskinJewishPopulation2012} As additional confirmation, we used ~\citeauthor{sheskinCollegeCampus2013}'s method to calculate an ``expansion factor'' based on the Pew 2013\footnote{We use the 2013 estimate rather than the 2020 estimate in order to align with Census data, of which 2010 is the latest available.} Jewish population estimate ~\cite{pew2013Report} in conjunction with the U.S.\ Census's 2010 surname frequency data ~\cite{us-census-list}.\footnote{ To do this, we counted the number of people with DJNs represented on the list, accounted for non-Jews with DJNs as described in \ref{methods:adt-data-processing}, and compared the resulting number to the total number of Jews in the U.S.\ counted by Pew.} We found that $\sim$9.15--11.18\% of the U.S.\ Jewish population at the time was covered by the DJN frame, which is aligned with prior estimates.

\subsubsection{Representativeness}
Another concern with using a DJN-only frame is as follows: making claims about all Jews on the basis of Jews with DJNs relies on the assumption that the latter group does not differ significantly in character from the former. A number of studies have been conducted on the representativeness of DJNs with mixed conclusions (in part due to lack of standardization with regards to which DJNs are used in a given study). It seems to be largely the case that DJN samples underrepresent intermarried Jews and their children, Jews with self-defined partial, mixed, or non-religions connections to Judaism, younger Jews, and (expectedly) Jews without Jewish parents ~\cite{cohenDeficientIfNotDistorted2016, dutwinEverythingToConsider2016, lazerwitzSomeCommentsOnDJNs1986, sheskinGoodPractices2016}. Critics of DJNs generally argue that the type of Jews least likely to be counted by DJNs are ``on the margin,'' which is especially problematic for studies whose goal it is to help Jewish community leaders best serve their constituents ~\cite{cohenDeficientIfNotDistorted2016}. However, because our study strictly estimates population size without surveying individuals, representation is only important insofar as it relates to likelihood of producing IP that appears in our dataset. As long as this is the case, our expansion factor should account for any undercount.

While we do not have prior reason to suspect such a bias, it is a limitation of our method that we cannot test for it directly. There is some evidence to suggest that DJN samples do not show significant differences in income ~\cite{cohenDeficientIfNotDistorted2016, sheskinGoodPractices2016} and education ~\cite{himmelfarbSamplingByEthnicSurnames} as compared to the general population; we did not find direct comparisons between the occupations of DJN and non-DJN samples (which would be the most direct proxy). Absent a compelling hypothesis for why DJN samples would significantly differ from the general population on occupation or writing output, it is reasonable to assume that our estimates provide order-of-magnitude bounds for the target variables.

\subsubsection{Broader Considerations}
While it is not the focus of this paper and we defer to the large literature on this topic for further discussion (e.g. ~\cite{black_ibm_2012}), it is useful to reflect on the broader considerations surrounding our need to rely on methodologies like those described above. Historical context is critical here. Previous databases of people with Jewish identity have been extremely dangerous to the Jewish community, with notable examples of these databases contributing to the deaths of millions of Jews in World War II ~\cite{black_ibm_2012}. Indeed, there is at least one person who was killed by the Nazi government for intentionally introducing noise into these databases ~\cite{noauthor_rene_2023}.

Given this history, and although this is out of scope for this work and not necessarily a consensus view, there is a reasonable belief that the best balance between being able to make data-driven assessments relevant to the Jewish community and protecting members of the community from serious material harm is through a noisy sensor like those provided by DJNs.

\subsection{Measuring IP Dispossession by Industry}
\label{app:ip-disp-by-industry}
Above, our analysis focused heavily on looking at carefully selected subsets of training data, each of which mapped to a particular job category or industry. 

Many LLMs have been trained using filtered subset of the wide-ranging Common Crawl dataset. This data is not as structured as any of the the specific Pile subsets we looked at, which each have underlying institutional norms that drive some of the implicit formatting, tone, and content standards.

Past work has already identified evidence of IP related to specific jobs in LLM training data. One case study showed that GPT-4 appears to have memorized content from the New York Times ~\cite{diakopoulosFindingEvidenceMemorized2023}, and the lawsuit against OpenAI filed by the New York Times provided further evidence of this memorization \cite{grynbaum_times_2023}. Investigation by the Washington Post and AI2 looked into the domains that contributed the most tokens to Google's C4 ~\cite{raffel2020exploring} dataset (one filtered version of Common Crawl) ~\cite{schaulSecretListWebsites}. In the top 10 domains alone, we see domains with IP produced by people in law (e.g. patents.google.com), media and journalism (e.g. nytimes.com), and science and medicine (e.g. 	journals.plos.org). The descriptive stats from ~\citeauthor{schaulSecretListWebsites}'s analysis~\cite{schaulSecretListWebsites} further substantiate the idea that IP-heavy job categories like law, media, journalism, science, and medicine help constitute much of the LLM training data outside careful subsets.

As a supplementary analysis, we conducted a small investigation into another filtered Common Crawl variant, RefinedWeb ~\cite{penedo2023refinedweb}. We approximated a random sample by randomly downloading 0.4\% of the roughly 5300 data files shared by the RefinedWeb curators on HuggingFace. We ranked the domains in this sample by number of total words. We found found very similar results to the C4 investigation -- journalists, scientists, medical researchers, lawyers, and other professional classes were dominant. Specifically, of the top 200 domains in C4, 131 were in at least the top 1000 of RefinedWeb. There are numerous subjective design choices involved in this minor analysis (when to consider subdomains like `patents.google.com', how to label a domain as pertaining to a specific job category), so we leave a full comparison along these lines to future work beyond the scope of our case study focusing on Jewish Americans.

The prior results related to job-specific memorization and potential IP dispossession further suggests that any groups whose economic well-being is highly tied to job market participation in IP heavy fields may be especially vulnerable to economic harms from the deployment of any labor-replacing LLM-based technologies.

We compared the self-reported job category numbers from a Pew Research study on Jewish Americans to U.S. Bureau of Labor Statistics numbers on the distribution of American workers by job category. While the BLS and job categories reported by Jewish respondents to the Pew survey did not map directly to each other, we were able to manually find some close mappings between them. This process required us to manually map the BLS categories to the Pew categories (which were not based on any formal taxonomy).

Based on this mapping, it seems likely that Jewish American workers participate in several IP-relevant professions at high rates, potentially around 4x for law and STEM. 

Thus, it is also likely possible to support the broad argument of our paper -- that Jewish Americans and other minority groups might be especially prone to IP dispossession and therefore to economic harms -- primarily using data on relative representation of group members in different job categories and connecting these job categories to LLM training data and LLM-replaceable labor.

In the future, it could make sense to investigate IP dispossession targeted at other groups with many members in these fields. Alternatively, professional organizations may wish to lead the charge themselves.

\end{document}